\documentclass[aps,pra,twocolumn,groupedaddress,showpacs,superscriptaddress,amssymb,amsmath]{revtex4-2}
\usepackage{natbib}
\usepackage[utf8]{inputenc}
\usepackage{graphicx}
\usepackage{tabularx}
\usepackage{xcolor}
\usepackage{amsmath}

\usepackage{comment}
\usepackage{dcolumn}
\usepackage{hyperref}
\definecolor{blue1}{HTML}{274057}
\hypersetup{
	colorlinks=true,
	linkcolor=blue,
	citecolor=blue,
    urlcolor=blue
}

\usepackage{bm}
\usepackage{epsf}
\usepackage{braket}
\usepackage{tensor}
\usepackage{soul}
\usepackage{mathrsfs}
\usepackage{mathtools}
\usepackage{multirow}

\usepackage{amsfonts}

\newcommand{\be}{\begin{equation}}
	\newcommand{\ee}{\end{equation}}
\newcommand{\bea}{\begin{eqnarray}}
	\newcommand{\eea}{\end{eqnarray}}

\def\la{\langle}

\def\ra{\rangle}

\newcommand{\sutdphys}{Science, Mathematics and Technology Cluster, Singapore
University of Technology and Design, 8 Somapah Road, 487372 Singapore}
\newcommand{\sutdepd}{EPD Pillar, Singapore University of Technology and Design, 8 Somapah Road, 487372 Singapore}

\newcommand{\cqt}{Centre for Quantum Technologies, National University of Singapore 117543, Singapore}

\makeatletter
\newsavebox{\@brx}
\newcommand{\llangle}[1][]{\savebox{\@brx}{\(\m@th{#1\langle}\)}%
	\mathopen{\copy\@brx\kern-0.5\wd\@brx\usebox{\@brx}}}
\newcommand{\rrangle}[1][]{\savebox{\@brx}{\(\m@th{#1\rangle}\)}%
	\mathclose{\copy\@brx\kern-0.5\wd\@brx\usebox{\@brx}}}
\makeatother

\begin{document}

\author{Katha Ganguly}
\email{katha.ganguly@students.iiserpune.ac.in}
\affiliation{Department of Physics,
		Indian Institute of Science Education and Research, Pune 411008, India}

\author{Dario Poletti} 
\email{dario\_poletti@sutd.edu.sg }
\affiliation{\sutdphys}
\affiliation{\sutdepd}
\affiliation{\cqt} 

\author{Bijay Kumar Agarwalla}
\email{bijay@iiserpune.ac.in}
\affiliation{Department of Physics,
		Indian Institute of Science Education and Research, Pune 411008, India}

\title{Long-time Dynamics of Many-body Open Quantum Systems using Quantum Generating Functions}

\date{\today}

\begin{abstract}
The interplay between coherent unitary evolution and environment-induced dissipation can give rise to a wide range of non-equilibrium dynamics in open quantum systems, ranging from interesting transport phenomena to dynamical phase transitions. However, accessing such long-time dynamics remains challenging for the existing methods developed for the simulation of many-body open quantum systems. 
We address this problem by developing a quantum generating function (QGF) formalism for open many-body systems
for both ensemble-averaged dynamics governed by a Markovian quantum master equation and trajectory-resolved dynamics described by the quantum trajectory formalism, including quantum jumps, and quantum state diffusion. Our approach computes the dynamics of higher-order moments and fluctuation statistics without explicitly evolving the quantum state, thereby providing an efficient and scalable approach for investigating many-body open quantum systems. We demonstrate the versatility of the formalism by applying it to both the integrable open XXZ chain and the nonintegrable open next-nearest-neighbor XXZ spin chain, where it uncovers distinct initial state dependent long-time transport regimes. Our work establishes quantum generating functions as an efficient and scalable framework for investigating long-time dynamics in open quantum many-body systems. 
\end{abstract}

  \maketitle  

{\it Introduction.--} 
Understanding the dynamics of many-body open quantum systems is a central challenge in modern condensed matter physics, both from theoretical~\cite{AMJFeb1982,TPSep2012,ACApr2017,SGJul2017,TGLDec2022,Sarang2025,MAP2025,PN2025,RS2025,Ray_2026} and experimental perspectives~\cite{RB2012,IB2012,AH2012,CM2019,MK2022,Immanuel2024,XZJan2025,CFRoos_2025}. 
These systems offer a rich playground of nonequilibrium phenomena with interesting long time dynamics and steady state properties which are the result of the interplay between coherent unitary dynamics and environment induced dissipation~\cite{Archak_2018,Fujimoto2022,Ganguly2024,Fazio2025}. 
This can result in a variety of transport regimes, e.g. ballistic, diffusive, subdiffusive, superdiffusive, and insulating \cite{TGLDec2022,APDec2021,MSMay2023,MSOct2023}, different thermodynamics functioning regimes \cite{Binder2018,GauthameshwarPoletti2025}, and dynamical transitions \cite{Kessler_2012_Dissipative, Fitzpatrick_2017_OpenExp}.      
However, understanding the long time behaviour of large open many-body systems has remained a central challenge due to the exponential growth of Hilbert space dimension. Over the years, several numerical techniques have been developed which have their own advantages and limitations, for example, tensor network based matrix product states (MPS) are useful to represent low entangled state, whereas numerics become expensive with increasing entanglement \cite{USJan2011, Daley_2014_Review}. On the other hand, neural network states can efficiently represent states with volume-law entanglement, but limited to short times or steady states \cite{Lange_2024_NQS_Review, Sinibaldi_2026_NeuralGalerkin, Schmitt_2025_tNQS, Hou_2026_SpacetimeTDSE, wang2026continuoustimeparametrizationneuralquantum, Hartmann_Carleo_2019, Vicentini_2019_Purified, Yoshioka_Hamazaki_2019, Luo_2022_Autoregressive, Reh_2021_AutoregressiveOpen, Vicentini_2022_GHDO, Zhang_2025_NQP}.

A recent breakthrough in the simulation of long-time dynamics in isolated quantum many-body systems has been achieved through the introduction of the quantum generating function (QGF) formalism~\cite{Prosen_QGF1}, which is based on full counting statistics of an observable and introduces a counting operator whose operator-space entanglement can remain comparatively low even at long times. The idea of  evolving the counting operator instead of states facilitates the computation of different moments of an observable over significantly long timescales.
This approach has recently been applied to isolated many-body systems to compute the statistics of integrated charge currents where various interesting long time dynamics is observed~\cite{Prosen_FV_2026,devendra_2026}.
However, an efficient generalization of the QGF formalism to open quantum many-body systems is still an open problem. Establishing such a framework would potentially provide access to a wide range of long-time dynamical phenomena, including transport, dynamical phase transitions, and other emergent nonequilibrium behaviors that have remained largely inaccessible to existing numerical approaches. Furthermore, studies of long-time quantum dynamics and transport in many-body systems have largely focused on initial states close to infinite temperature or highly polarized states, such as domain-wall configurations. 
In contrast, our work investigates dynamical transport in a fundamentally different manner, which broadens significantly the types of initial conditions that can be considered.    

In this work, we show how one can study the long-time dynamics of many-body open quantum systems in regimes where the conventional state-evolution methods by various numerical techniques become computationally impractical. This relies on extending the methods based on QGF to open quantum systems. 
Importantly, Markovian open quantum dynamics in the Gorini-Kossakowski-Sudarshan-Lindblad (GKSL) form, \cite{VG1976,L1976}, is commonly described within two complementary frameworks~\cite{BPOQS,carmichael2009,IROct2015,dutta2025}: (i) Ensemble-averaged description governed by the  GKSL quantum master equation (QME) ~\cite{L1976,VG1976}, and (ii)  trajectory-resolved description provided by quantum trajectory formalism, whether with quantum jump or with a state diffusion equation~\cite{MBPJan1998,GISIN1992315,ToddBrun2000,carmichael2009,Molmer1992}. Each approach can allow to better study different quantum systems, depending on the specific features of the setup. Here, we show how to formulate the QGF approach within both frameworks.  

{\it QGF Procedure for open systems.--} 
The time evolution of the system is governed by the Liouvillian superoperator $\mathcal{L}$ according to
${d\rho/dt}=\mathcal{L}\rho$,
where $\mathcal{L}$ comprises both the unitary contribution generated by the Hamiltonian and the dissipative contribution arising from environmental interactions. 
Within the QGF framework, one can compute either the moments of the observable $\hat{Q}(t)$ itself (single time measurement protocol) or that of the net transferred quantity $\Delta\hat{Q} :=\hat{Q}(t)-\hat{Q}(0)$ (two time measurement protocol).  
To compute the QGF for $\Delta\hat{Q}$, 
we employ the two-time measurement protocol--measuring the state at initial time $t=0$ and at time $t$. Therefore, starting from the the initial state $\rho_0$, a projective measurement of the observable $\hat{Q}$ is performed which leads to the post measurement state $\widetilde{\rho}_0=\sum_{Q_0}\hat{P}_{Q_0}\rho_0 \hat{P}_{Q_0}$, where $Q_0$ is an eigenvalue of the operator $\hat{Q}$. The state $\widetilde{\rho}_0$ is then evolved by the effect of Liouvillian $\mathcal{L}$ followed by a second projective measurement at time $t$ which gives the final post-measurement state $\widetilde{\rho}_t=\sum_{Q_t}\hat{P}_{Q_t}e^{\mathcal{L}t}[\widetilde{\rho}_0] \hat{P}_{Q_t}$. Hence, the probability distribution of the net transferred charge $\Delta Q$ during the time interval $[0,t]$ is given by~\cite{Bijay2012},
\begin{equation}
\!\!P_t(\Delta Q) =\!\!\sum_{Q_t,Q_0}\!\!\delta(Q_t\!-\!Q_0\!-\!\Delta Q)\,\mathrm{Tr}\Big[\hat{P}_{Q_t}e^{\mathcal{L}t}[\hat{P}_{Q_0}\rho_0 \hat{P}_{Q_0}]\Big]. \label{eq:Prob_dist_delQ}
\end{equation}
The QGF $G(\lambda,t)$ is obtained by performing a Fourier transformation of $P_t(\Delta Q)$ with respect to the counting field $\lambda$. It is given as
\begin{align}
G(\lambda,t)=\mathrm{Tr}\Big[e^{\mathcal{L}^{\dagger}t}\big[\hat{R}(\lambda,0)\big]\hat{R}^{\dagger}(\lambda,0)\widetilde{\rho}_0\Big], \label{eq:G_lambda_def}
\end{align}
where $\hat{R}(\lambda,0)=e^{i\lambda \hat{Q}}$, which we refer to as the counting operator. To obtain Eq.~\eqref{eq:G_lambda_def}, we have replaced the Dirac delta function by $\delta(x)=(1/2\pi)\int_{-\infty}^{+\infty}dk \,e^{ikx}$ and used the relation $\mathrm{Tr}\big[Ae^{\mathcal{L}t}[B]\big]=\mathrm{Tr}\big[e^{\mathcal{L}^{\dagger}t}[A]B\big]$ where $\mathcal{L}^{\dagger}$ is the adjoint Liouvillian. Interestingly, for $\lambda=0$, 
$e^{\mathcal{L}^{\dagger}t}[\hat{R}(0,0)]= e^{\mathcal{L}^{\dagger}t}\, [\, \mathbb I\, ]
= \mathbb
I$ for all times $t$ which is a trivial matrix product operator (MPO). This follows from the fact that the identity operator is the right eigenvector of $\mathcal{L}^{\dagger}$ with eigenvalue $0$, a property referred to as the {\it unitality} of $\mathcal{L}^{\dagger}$. 
Consequently, for small values of $\lambda$, the counting operator $\hat{R}(\lambda,t)=e^{\mathcal{L}^{\dagger}t}[\hat{R}(\lambda,0)]$ can remain close to the identity operator up to a significantly large time window. As a result, the bond dimension of $\hat{R}(\lambda,t)$ remains low throughout the dynamics and can always be controlled by appropriately tuning $\lambda$. However, choosing $\lambda$ to be too small is also not optimal, as it requires a larger bond dimension to accurately capture the dynamics~\cite{supp_mat}. This controllability of the bond dimension makes the simulation of $\hat{R}(\lambda,t)$ particularly advantageous for simulating long-time open many-body dynamics, especially in comparison with conventional state evolution using MPS, where the bond dimension can, in principle, grow uncontrollably with time.

In this work, we use the Time Evolution Block Decimation (TEBD) algorithm to simulate the time dynamics of $\hat{R}(\lambda,t)$ and use that to compute the QGF $G(\lambda,t)$ at each time instant. 
From the QGF, we analytically continue $\lambda$ to the complex plane  $\lambda=|\lambda|e^{i\phi}$ and expand $G(\lambda,t)$ in orders of $\lambda$   
as~\cite{Prosen_QGF1},
\begin{align}
G_\phi(\lambda,t)=1+\sum_{n=1}^{\infty}\frac{(i|\lambda|)^n}{n!}e^{in\phi}\mu_n(t). \label{eq:G_expanded}
\end{align}
This complex plane extension helps to obtain different moments appropriately by choosing proper real and imaginary part of $G_\phi(\lambda,t)$ for a given $\phi$ value. For example, the first and second moment of $\Delta \hat{Q}$ are given by 
\begin{align}
    &\mu_1(t)=\frac{1}{|\lambda|}\mathrm{Im} \big[G_0(|\lambda|,t)\big] + O(\lambda^2),\label{eq:mu_1}\\
    &\mu_2(t) = \frac{2}{|\lambda|^2}\big(1-\mathrm{Re}[G_0(|\lambda|,t)]\big)+ O(\lambda^2),\label{eq:mu_2}
\end{align} 
\begin{figure*}
    \centering
\includegraphics[width=0.3\linewidth]{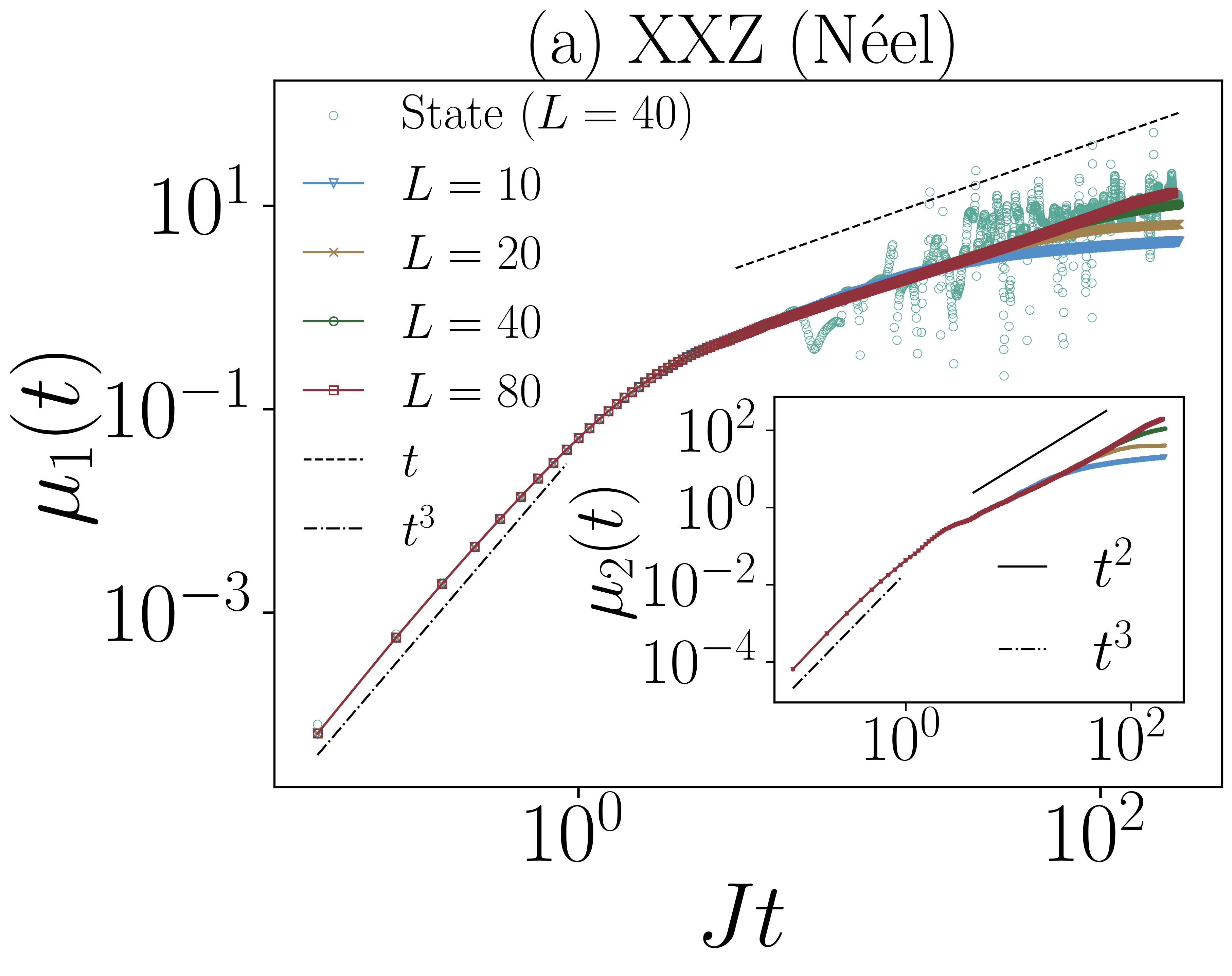}%
\includegraphics[width=0.3\linewidth]{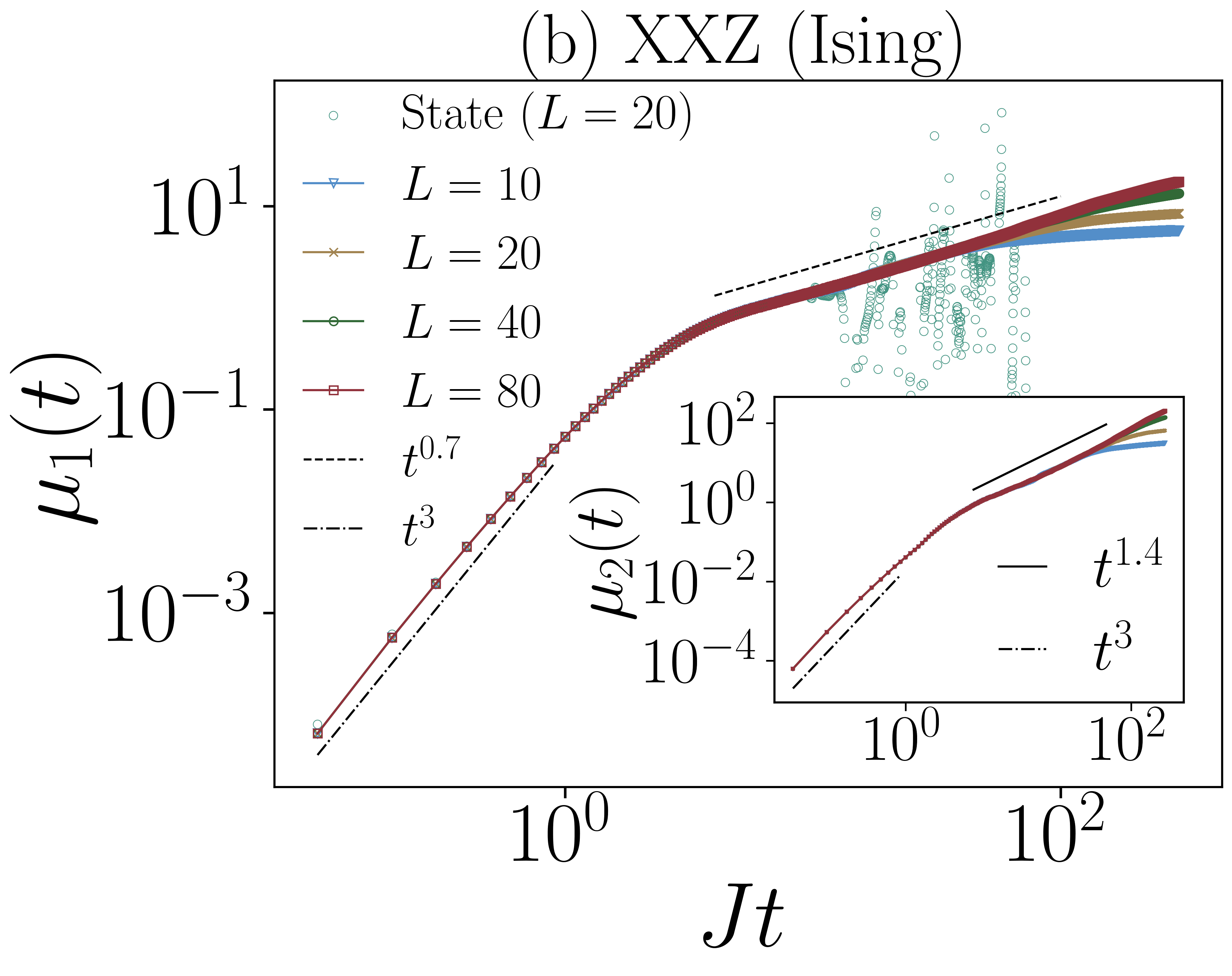}%
\includegraphics[width=0.3\linewidth]{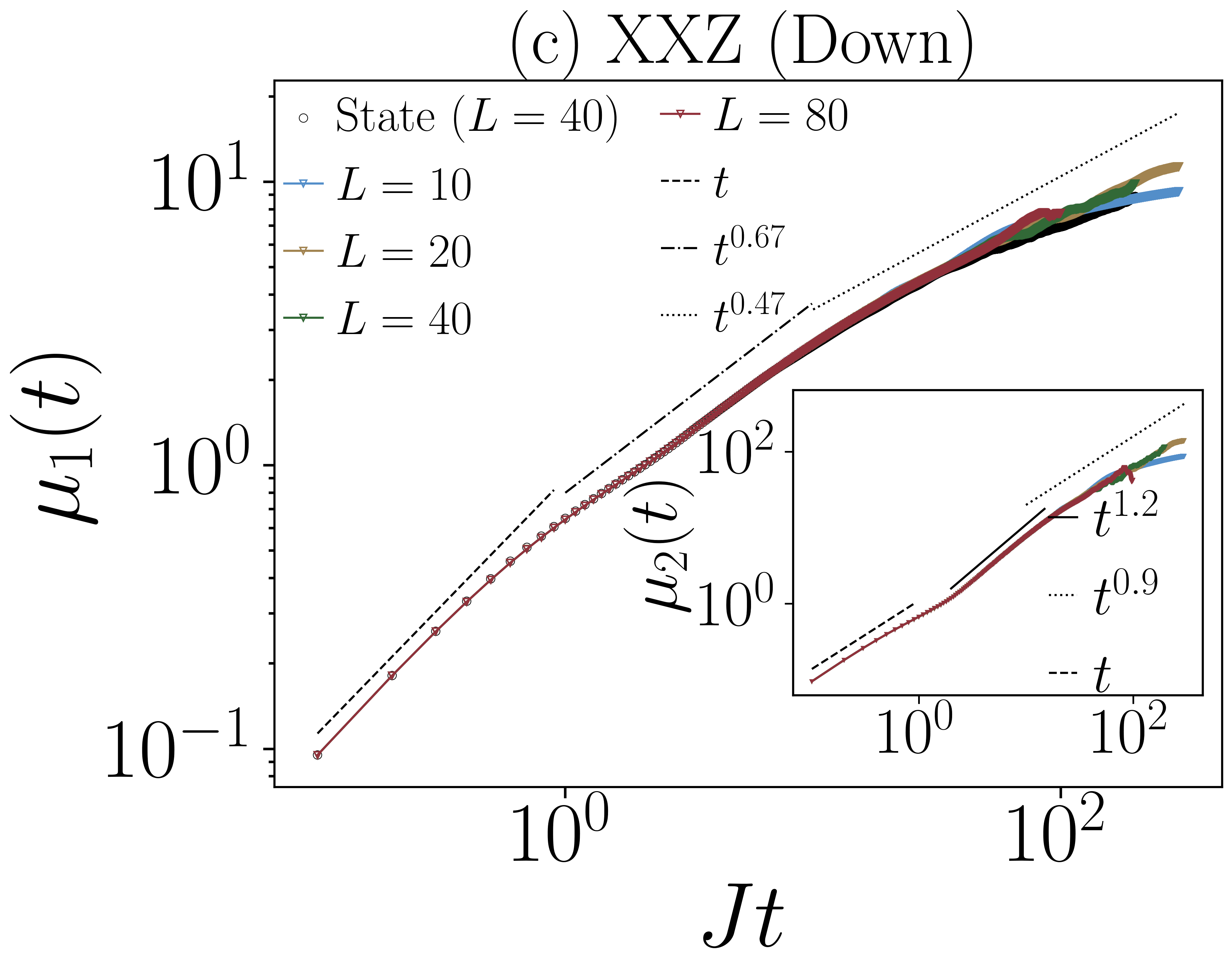}
\includegraphics[width=0.3\linewidth]{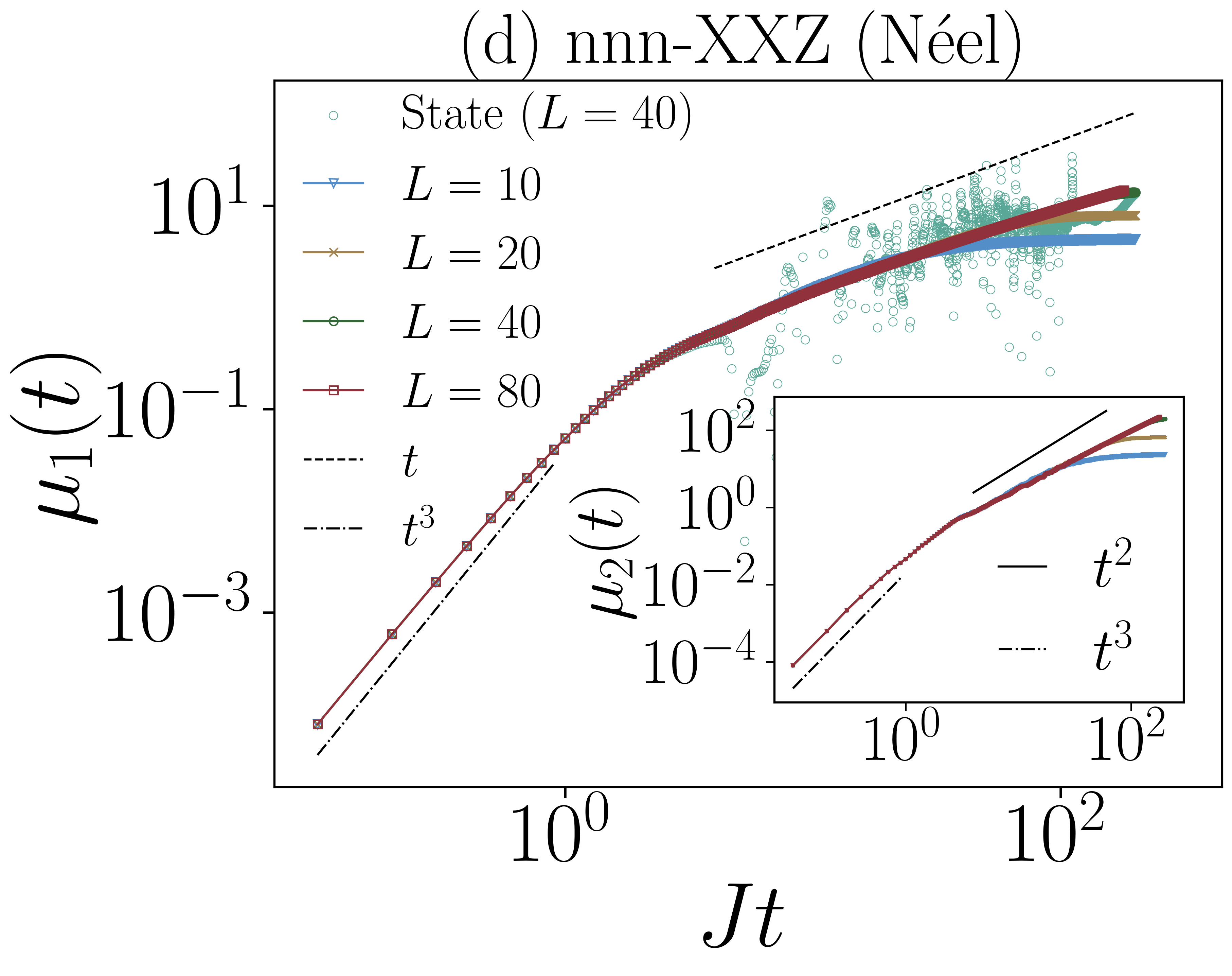}%
\includegraphics[width=0.3\linewidth]{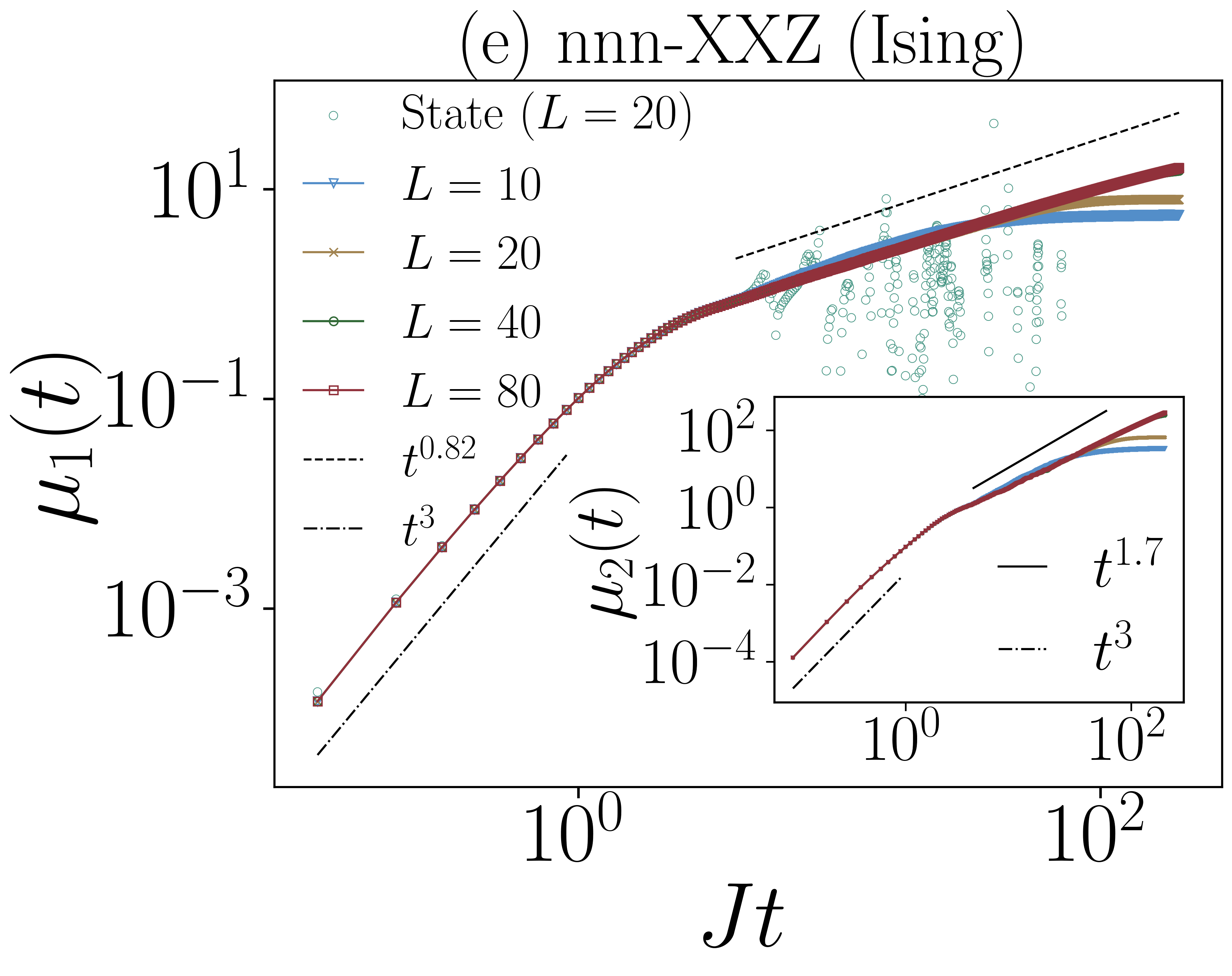}%
 \includegraphics[width=0.3\linewidth]{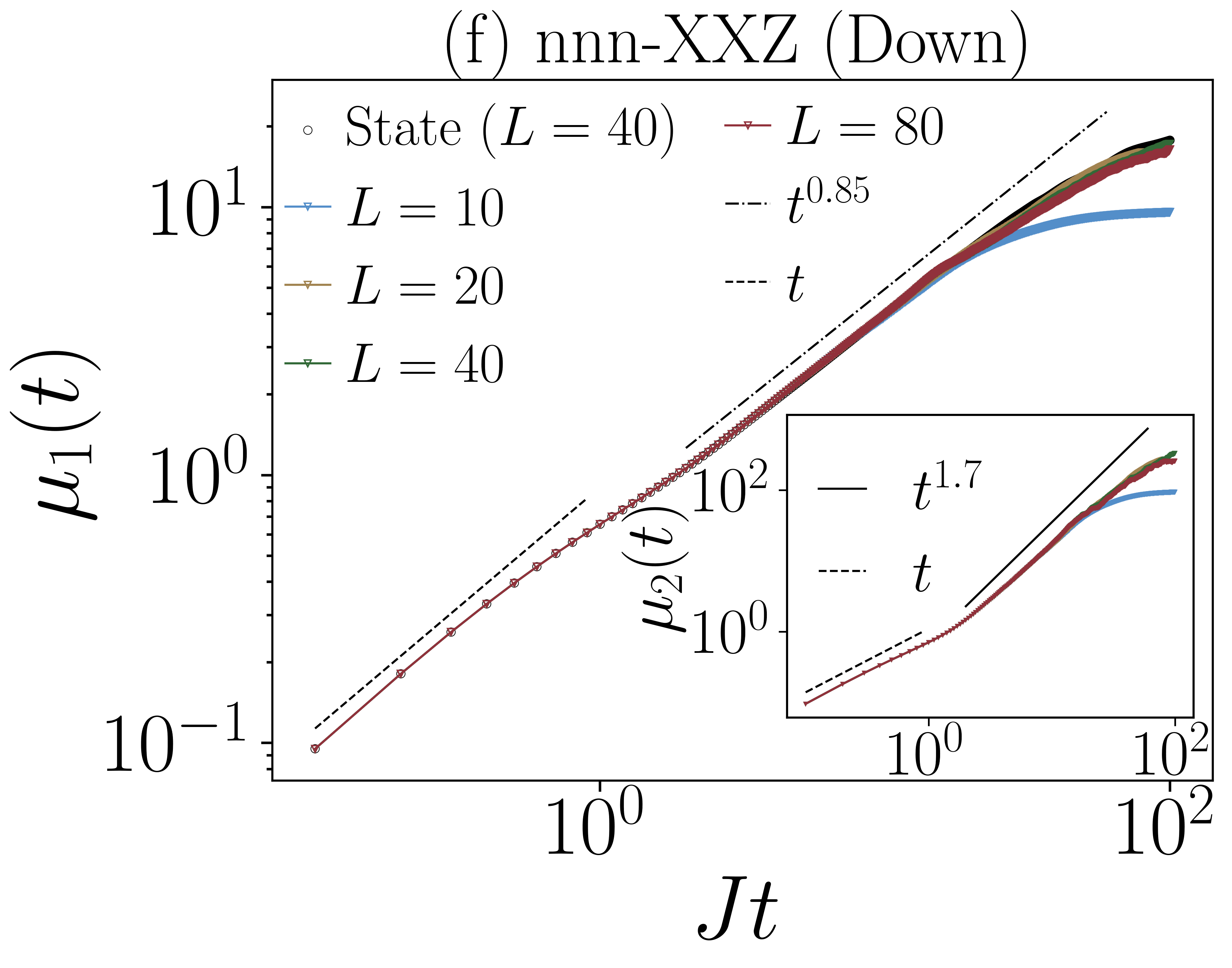}%
    \caption{First moment $\mu_1(t)$ (main figures) and second moment $\mu_2(t)$ (inset figures) of the transferred magnetization for the XXZ chain (top panel) and nnn-XXZ chain (bottom panel), subjected to excitation injection for the initial states (a) N\'eel, (b) Ising-type, and (c) all spins down. Results are shown for system sizes $L=10,20,40,$ and $80$. The scattered data points correspond to the state-evolution method using MPS, while the solid curves are obtained from the QGF approach. For both models, we set $\Delta=1.0J$, $\Gamma_G=1.0J$, $|\lambda|=0.03$, and $\phi=0$. For the nnn-XXZ model, the next-nearest-neighbour coupling is $J_b=J$. In both the QGF and MPS simulations, the dynamics is simulated using the TEBD algorithm with a second-order Suzuki-Trotter decomposition, time step $dt=0.1$, maximum bond dimension $256$, and truncation cutoff $10^{-10}$.}
    \label{fig:setup_I}
\end{figure*}
where we choose $\phi=0$. To obtain the higher moments $(n \geq 3)$, one may need to choose much smaller value of $|\lambda|$ and an appropriate phase $\phi$~\cite{Prosen_QGF1}. Following this recipe, one can analyze the long time dynamics of different moments in open many body systems.

{\it Long time dynamics of many-body quantum systems with a localized particle injection.--} We now apply the developed QGF framework to an open many-body spin-$1/2$ lattice system governed by the QME
\begin{equation}
\frac{d\rho}{dt}=\mathcal{L}\rho=-i[\hat{H},\rho]+\sum_j\Big(\hat{L}_j\rho \hat{L}_j^\dagger-\frac{1}{2}\{\hat{L}_j^\dagger \hat{L}_j,\rho\}\Big).
\label{eq:GKSL}
\end{equation}
Here, the Hamiltonian $H$ is given by
\begin{align}
    \hat{H}=J\sum_{j=1}^{L-1}&\big[\hat{S}^x_j\hat{S}^x_{j+1}+\hat{S}^y_j\hat{S}^y_{j+1}+\Delta \hat{S}_j^z\hat{S}_{j+1}^z\big]\nonumber\\&+J_b\sum_{j=1}^{L-2}\big[\hat{S}^x_j\hat{S}^x_{j+2}+\hat{S}^y_j\hat{S}^y_{j+2}+ \hat{S}_j^z\hat{S}_{j+2}^z\big]. \label{eq:Ham}
\end{align}
For $J_b=0$, Eq.~\eqref{eq:Ham} reduces to the integrable XXZ model, whereas $J_b\neq 0$ corresponds to the non-integrable next-nearest-neighbour XXZ (nnn-XXZ) model.
The dissipative dynamics is encoded in the jump operators $\hat{L}_j$, appearing in the second and third terms of Eq.~\eqref{eq:GKSL}. For these models, we focus on the quantum dynamics of the moments of transferred magnetization operator $\Delta \hat{S}^z$ where $\hat{S}^z=\sum_{j =1}^L{S}^z_j$, when the lattice is subjected to a single dissipative channel with jump operator $\hat{L}=\sqrt{\Gamma_G} \hat{S}_1^{+}$, corresponding to a local injection of spin excitations at the left boundary of the chain, thereby driving it out of equilibrium. Here $\Gamma_G$ is the injection rate and $\hat{S}_1^{+}$ is the spin-excitation operator acting on the first lattice site. 
We consider three different initial states for both XXZ and nnn-XXZ models: (a) the N\'eel state, $|\psi_0\rangle=|\!\!\uparrow\downarrow\cdots\uparrow\downarrow\rangle$; (b) an Ising-type state~\cite{Jiang_2026}, $|\psi_0\rangle=|m_1\rangle\cdots|m_L\rangle$, where $|m_j\rangle=|\!\!\uparrow\rangle$ if $\cos(\pi j/3+\pi/6)>0$ and $|m_j\rangle=|\!\!\downarrow\rangle$ otherwise; and (c) a fully polarized state with all spins down, $|\psi_0\rangle=|\!\!\downarrow\downarrow\cdots\downarrow\rangle$. We investigate quantum dynamics for the first and second moment and observe interesting initial condition dependent long-time dynamics.
Open quantum dynamics study for such a setup over a large time-window is significantly limited. For example, performing the conventional state evolution via MPS under such a scenario can result in an uncontrolled growth of the bond dimension. The developed QGF method, instead, can provide access to long time dynamics and hence can potentially showcase different interesting dynamical transport regimes. The different transport regimes can be distinguished by the dynamical scaling of the moments $\mu_n(t)$ of the transferred charge with time. In particular, ballistic transport is characterized by $\mu_n(t)\propto t^n$, whereas diffusive transport exhibits the scaling $\mu_n(t)\propto t^{n/2}$. Superdiffusive and subdiffusive transport are characterized by $\mu_n(t)\propto t^{n\nu}$, with $1/2<\nu<1$ and $\nu<1/2$, respectively.

In Fig.~\ref{fig:setup_I}, we present the results of the long time dynamics for the three different initial states discussed previously. The open quantum dynamics is simulated using the TEBD algorithm by evolving the initial counting operator $\hat{R}(\lambda,0)=e^{i\lambda \hat{S}^z}$ under the action of adjoint Liouvillian $\mathcal{L}^{\dagger}$ which is given by,
\begin{equation}
    \mathcal{L}^{\dagger}[\hat{R}]=i[\hat{H},\hat{R}]+ \Gamma_G \Big(\hat{S}_1^-\, \hat{R} \, \hat{S}_1^+-\frac{1}{2}\{\hat{S}_1^- \hat{S}_1^+,\hat{R}\}\Big).
\end{equation}
Here we employ a second-order Suzuki-Trotter decomposition of $e^{\mathcal{L}^{\dagger}t}$ with a time step $dt=0.1$, and set $|\lambda|=0.03$ and $\phi=0$.
In Fig.~\ref{fig:setup_I}(a)-(c), we plot the first and second moments, $\mu_1(t)$ and $\mu_2(t)$, for the XXZ chain with anisotropy parameter $\Delta=1.0J$ starting from the N\'eel state [Fig.~\ref{fig:setup_I}(a)], the Ising-type state [Fig.~\ref{fig:setup_I}(b)], and the fully polarized down state [Fig.~\ref{fig:setup_I}(c)]. We find a striking dependence of the long-time transport dynamics on the initial state. For the N\'eel state, both $\mu_1(t)$ and $\mu_2(t)$ exhibit an initial growth proportional to $t^3$ at very early times $Jt\ll 1$, followed by a ballistic transport regime characterized by $\mu_1(t)\propto t$ and $\mu_2(t)\propto t^2$. In contrast, for the Ising-type initial state, the initial $t^3$ growth crosses over to a superdiffusive regime at late times, with $\mu_1(t)\propto t^{0.7}$ and $\mu_2(t)\propto t^{1.4}$. For the fully polarized down state, the early-time behavior is qualitatively different: both moments initially grow linearly in time rather than  $t^3$. This is followed by a transient superdiffusive regime, with $\mu_1(t)\propto t^{0.67}$ and $\mu_2(t)\propto t^{1.2}$, before crossing over to diffusive behavior at late times, characterized by $\mu_1(t)\propto \sqrt{t}$ and $\mu_2(t)\propto t$. These results demonstrate that the nonequilibrium transport dynamics of the open XXZ chain can depend sensitively on the initial state, giving rise to a plethora of transient and asymptotic transport regimes. These intriguing dynamical regimes warrant a more thorough theoretical investigation.
In contrast, the results for the nnn-XXZ chain presented in Fig.~\ref{fig:setup_I}(d)-(f) are qualitatively similar for all three initial states, i.e., $\mu_1(t)\propto t^{0.85}$ and $\mu_2(t)\propto t^{1.7}$. This insensitivity to the choice of initial state can be attributed to the chaotic nature of the underlying Hamiltonian. To highlight the limitations of conventional state evolution, we further plot $\mu_1(t)$ obtained from direct state evolution (scattered plots in Fig.~\ref{fig:setup_I}), where the density matrix $\rho$ is evolved under the GKSL equation. We find that state evolution rapidly becomes inaccurate at long times for both the N\'eel and Ising-type initial states, whereas it remains reliable for the fully polarized down state. The QGF therefore provides access to long-time transport dynamics revealing initial state dependent rich transport phenomena that are otherwise inaccessible through conventional state evolution. We further investigate the dependence of $G(\lambda,t)$ on $\lambda$ at different times and extract the probability distribution of the injected charge via inverse Fourier transform~\cite{supp_mat}. 

In the Supplemental Material, we further apply the developed QGF framework to a different open many-body setup, where the XXZ and nnn-XXZ chains are subjected to \textit{weak} local dephasing at every lattice site. Owing to the \textit{weak} dephasing, the conventional state-evolution approach once again becomes computationally intractable at long times, making the QGF framework necessary for investigating the long-time dynamics.

{\it QGF approach for quantum trajectories.--} 
An alternative and widely employed approach to study open quantum systems is the quantum trajectories formalism. In this approach, instead of evolving the mixed quantum state $\rho$, one evolves pure-state trajectories in a stochastic manner by unraveling the QME. As we will show later, this approach naturally gives insight into the statistics of the measurements. 
Furthermore, this approach offers a significant numerical advantage because a pure state is represented by a $2^L$-dimensional vector in Hilbert space, whereas a vectorized mixed state is represented by a $4^L$-dimensional vector in Liouville space. To accurately recover the open-system dynamics, however, averaging over sufficiently large ensemble of stochastic trajectories is required. Since individual trajectories are statistically independent, the required ensemble averaging can be performed efficiently using parallel computation. Beyond this computational advantage, individual quantum trajectories have been experimentally accessed, for example through continuous monitoring~\cite{wiseman2010,jacobs2014,Jacobs01092006} across a variety of quantum platforms~\cite{Rainer1986,Vijay2011,Kater2013}. It has thus enabled the observation and investigation of a variety of emergent phenomena, such as measurement-induced phase transitions~\cite{Skinner2019,Ashida2020,sebastian_2021}, measurement-induced symmetry restoration~\cite{Xhek2025,ganguly2026}, and charge-sharpening transitions~\cite{Sarang_2022,Feng_2025}.

Motivated by these developments, we extend the QGF approach to individual quantum trajectories to avoid evolving the pure quantum states explicitly. 
We can write the evolution of a stochastic trajectory as $|\psi_{t+dt}\ra=e^{\hat{K}^t_{dt}(\xi)}|\psi_t\ra$, where we refer $\hat{K}^t_{dt}(\xi)$ as the generator of the evolution for the quantum trajectory from time $t$ to $t+dt$. Note that in general, the generator $\hat{K}^t_{dt}(\xi)$ can depend on the instantaneous state $|\psi_t\ra$. For a single quantum trajectory, we then define a trajectory-resolved quantum generating function (QGF) as,
\begin{align}
G_\xi(\lambda,t)=\mathrm{Tr}\Big[e^{\hat{K}_{\xi}^{\dagger}}\big[\hat{R}(\lambda,0)\big]e^{\hat{K}_\xi}\hat{R}^{\dagger}(\lambda,0)\tilde{\rho}_0\Big], \label{eq:qgf_qj}
\end{align}
where, $e^{\hat{K}_\xi}=e^{\hat{K}_{dt}^t(\xi_t)}\,e^{\hat{K}_{dt}^{t-dt}(\xi_{t-dt})}\dots e^{\hat{K}_{dt}^{dt}(\xi_{dt})}$  represents the evolution operator evolving a single trajectory. Consequently, the counting operator $\hat{R}(\lambda,0)$ evolves as
$\hat{R}(\lambda,t)=e^{\hat{K}_\xi^{\dagger}}\hat{R}(\lambda,0)e^{\hat{K}_\xi}$. For notational simplicity, we suppress the explicit dependence on $\xi$ hereafter and write $\hat{K}_{dt}^t$ instead of $\hat{K}_{dt}^{t}(\xi_t)$. Similar to the ensemble-averaged evolution governed by the GKSL master equation, the trajectory propagator also satisfies the unitality property i.e., $e^{\hat{K}_{dt}^{\dagger}}e^{\hat{K}_{dt}}=\mathbb{\hat{I}}+O(dt^2)$ at every time step. Therefore,
for sufficiently small $\lambda$, $\hat{R}(\lambda,t)$ remains close to the identity operator, making its propagation computationally efficient.
\begin{figure}[h!]
    \centering
    \includegraphics[width=0.5\linewidth]{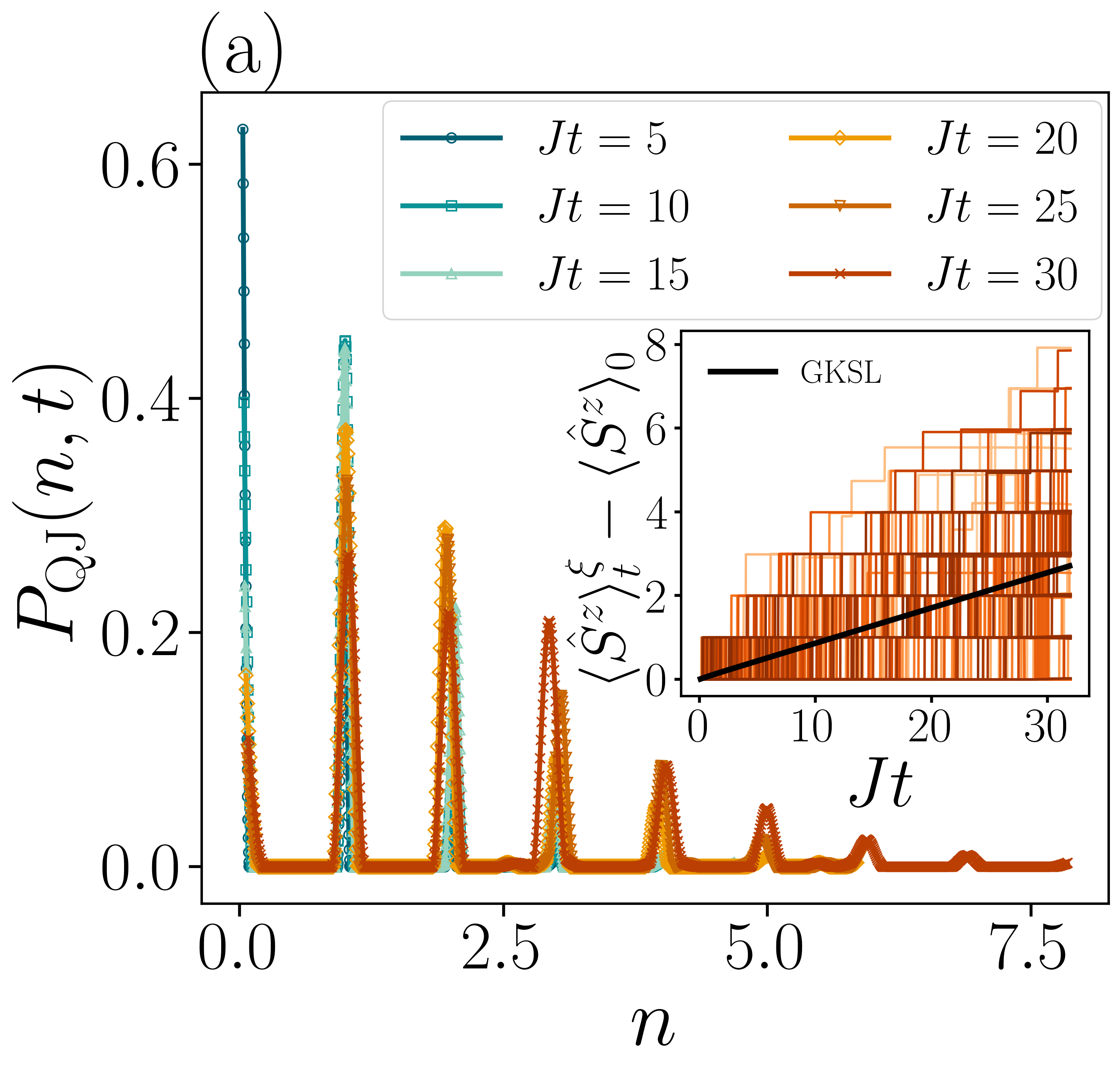}%
    \includegraphics[width=0.5\linewidth]{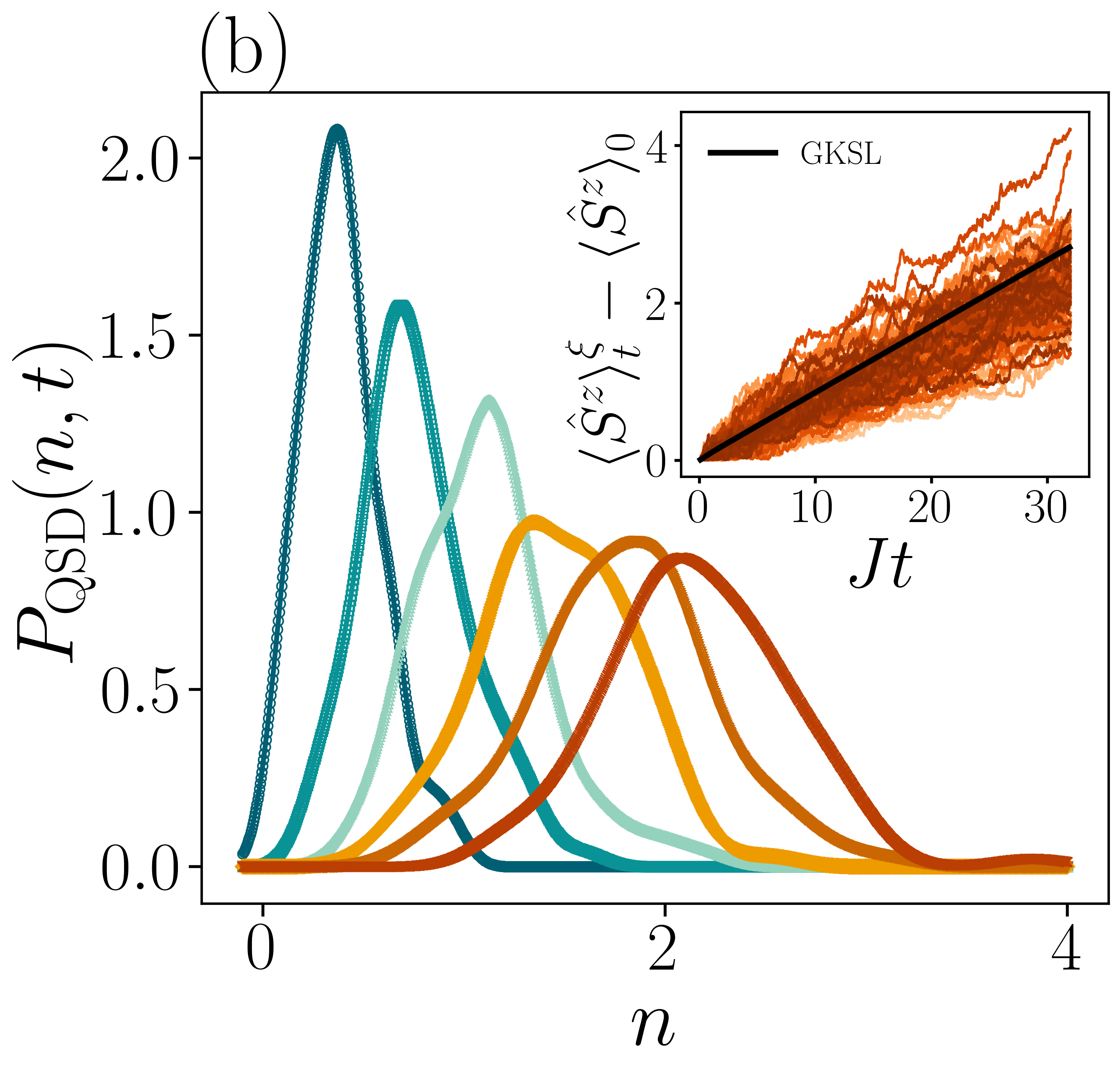}
    \caption{Probability distribution of the injected charge $\langle{S}^z\rangle_t^{\xi}-\langle{S}^z\rangle_0=n$, at different times $Jt$, obtained by sampling $200$ quantum trajectories for the (a) quantum-jump (QJ) and (b) quantum-state-diffusion (QSD) protocols. In both cases, we consider the XXZ chain ($J_b=0$) with anisotropy $\Delta=1.0J$. The remaining parameters are $\Gamma_G=0.1J$, and $\lambda=0.03$, for a system of size $L=20$. The simulations are performed using TEBD algorithm with bond dimension $\chi=256$ and a truncation cutoff $10^{-10}$. 
    The insets in (a) and (b) show the time evolution of the injected charge $\langle{\hat{S}^z}\rangle_t^{\xi} - \langle{\hat{S}^z}\rangle_0$ for different trajectories. The solid black curve in the inset denotes the ensemble-averaged injected charge, which coincides with the result obtained from the GKSL master equation.}
    \label{fig:traj}
\end{figure}
In what follows, we discuss two different kinds of stochastic trajectory protocols, namely the quantum jump (QJ), and the quantum state diffusion (QSD) for the XXZ setup with locally injected spin excitation, as discussed before.
For the (QJ) protocol~\cite{Molmer1992,Landi_FCS_traj_2024} the quantum state evolves via stochastic Schr\"odinger equation (SSE) given by,
\begin{align}
    d|\psi_t\rangle = -i\hat{H}_{\rm eff}\,dt|\psi_t\ra+dN_t\left(\frac{\hat{S}_1^{+}}{\sqrt{\la \hat{S}_1^{-}\hat{S}_1^{+}\ra}}-1\right)|\psi_t\ra,\label{eq:qj_sse}
\end{align}
where $\hat{H}_{\rm eff}=\hat{H}-i\,\Gamma_G\big(\hat{S}_1^{-}\hat{S}_1^{+}-\la \hat{S}_1^{-}\hat{S}_1^{+}\ra_t\big)/2$. $dN_t$ is a Poisson stochastic variable that takes value $dN_t\!=\!1$ with probability $\Gamma_G \, dt \, \la \hat{S}_1^{-}\hat{S}_1^{+}\ra_t$ and $dN_t=0$ with probability $1-\Gamma_G \, dt \, \la \hat{S}_1^{-}\hat{S}_1^{+}\ra_t$.  
For this protocol, we compute the total injected charge along each quantum-jump trajectory using the QGF in Eq.~\eqref{eq:qgf_qj} starting from the N\'eel state for XXZ chain with $\Delta=1.0J$. It is obtained as,
\begin{equation}
\mu_1(t)=\langle{\hat{S}^z}\rangle_t^{\xi}-\langle{\hat{S}^z}\rangle_0
=-\frac{1}{\lambda}\,\mathrm{Im}\!\left[G_\xi(\lambda,t)\right]
+O(\lambda^2),
\end{equation} 
where we choose $\lambda$ to be real by setting $\phi=0$. Here, we are interested in the distribution of the injected charge by sampling a large number of quantum trajectories.
Interestingly, for the QJ protocol, if one samples a sufficiently large number of quantum trajectories and computes the total injected charge $\langle{\hat{S}^z}\rangle_t^{\xi}-\langle{\hat{S}^z}\ra_0$ for each trajectory, and finally constructs its distribution $P_{\rm QJ}(n,t)$ (here $n=\langle{\hat{S}^z}\rangle_t^{\xi}-\langle{\hat{S}^z}\ra_0$), the resulting distribution is exactly identical to $P(n,t)$  obtained from the Fourier transform of the ensemble-averaged generating function $G(\lambda,t)$ under the GKSL master equation~\cite{Landi_FCS_traj_2024}~\cite{supp_mat}. 
We sample an ensemble of $200$ quantum trajectories to obtain the probability distribution $P_{\rm QJ}(n,t)$, shown in Fig.~\ref{fig:traj}(a) at different time snapshots. Owing to the discrete nature of quantum jumps generated by $\hat{S}_1^{+}$, the injected charge is always quantized in integer units, increasing by one with each jump event. The inset of Fig.~\ref{fig:traj}(a) shows the time evolution of the injected charge for $200$ trajectories. The staircase-like evolution, consisting of unit-height jumps separated by plateaus, clearly distinguishes the jump events from the no-jump evolution which is governed by $\hat{H}_{\rm eff}$.

For the same setup, we next implement the quantum state diffusion (QSD) protocol~\cite{NGisin_1992} which is another type of quantum trajectory and corresponds to a different continuous and weak measurement scheme. The experimental realization of such a process is possible through homodyne detection experiments~\cite{Yuen_83,Warszawski_2002,Landi_FCS_traj_2024}. The evolution of the pure state under the QSD is given by,
\begin{align}
&    d|\psi_t\ra = \Big[-i\hat{H}dt+d\xi_t\big(\hat{S}_1^{+}-\la \hat{S}_1^{+}\ra_t\big)\nonumber\\
   &\,\,+\frac{\Gamma_Gdt}{2}\big(\hat{S}_1^{-}\hat{S}_1^{+}-2\la \hat{S}_1^{-}\ra_t \hat{S}_1^{+}+\la \hat{S}_1^{-}\ra_t\la \hat{S}_1^{+}\ra_t\big)\Big]|\psi_t\ra. \label{eq:sse_qsd}
\end{align}
Here, $d\xi_t$ is Gaussian noise with mean $0$ and variance $\Gamma_G \, dt$ describing a Wiener process. Once again, we sample $200$ quantum trajectories and obtain the injected charge $\langle{\hat{S}^z}\rangle_t^{\xi}-\langle{\hat{S}^z}\ra_0$ for each trajectories using QGF in Eq.~\eqref{eq:qj_sse}. We plot the distribution of injected charge in the diffusive trajectories $P_{\rm QSD}(n,t)$ for different time snapshots in Fig.~\ref{fig:traj}(b), and the evolution of different trajectories in its inset. The distribution is Gaussian due to the underlying Wiener process. Moreover, unlike the QJ scenario, here the injected charge can take non-integer values as well, which makes this distribution significantly different from $P_{\rm QJ}(n,t)$. Overall, 
the average injected charge remains identical for both the QJ and QSD protocols.

\textit{Summary.--} In this work, 
we have developed a QGF formalism for 
Markovian open quantum systems. We have formulated the QGF framework for both the ensemble-averaged GKSL quantum master equation and stochastic quantum trajectories, each giving different insights into the dynamics, and each having different computational advantages. This generalization is fundamentally enabled by the {\it unitality} of the adjoint Liouvillian superoperator for the GKSL dynamics and of the trajectory propagator for quantum trajectories, both of which admit the identity operator as an eigenoperator.

We have demonstrated the versatility of the proposed framework in open quantum many-body systems, where conventional state-evolution methods become computationally extensive due to the rapid growth of operator-space or trajectory entanglement. 
In particular, we have studied the integrable XXZ and next-nearest-neighbor XXZ chains subjected to boundary spin injection and shown that the QGF formalism efficiently computes the higher-order moments of magnetization transport at long times. This allowed us to uncover a intriguing initial state dependence on the long-time transport regimes for the XXZ model.  
Furthermore, we have generalized the framework to both quantum-jump and quantum-state-diffusion unravelings by introducing a trajectory-resolved QGF, in this way allowing the simulate different types of measurement scenarios. 

Overall, this work establishes the QGF formalism as a computationally efficient framework for probing long time dynamics in many-body open quantum systems, particularly in regimes where conventional state-based approaches become intractable. Promising future directions include extending the framework to non-Markovian open-system dynamics and strong system-environment coupling regimes.

\vspace{0.1cm}
\textit{Acknowledgements.--}  
BKA acknowledges the CRG grant No. CRG/2023/003377 from ANRF, Government of India. KG would like to acknowledge the Prime Minister's Research Fellowship (ID- 0703043), Government of India for funding. KG and BKA acknowledge the National Supercomputing Mission (NSM) for providing computing resources of ‘PARAM Brahma’ at IISER Pune, which is implemented by C-DAC and supported by the Ministry of Electronics and Information Technology (MeitY) and DST, Government of India. 
DP acknowledges support from the Singapore Ministry of Education from grant MOE-T2EP50123-0017. 

\bibliography{references}

@article{Prosen_QGF1,
  title = {Efficient Computation of Cumulant Evolution and Full Counting Statistics: Application to Infinite Temperature Quantum Spin Chains},
  author = {Valli, Angelo and Moca, C. P and Werner, M. A. and Kormos, M. and Krajnik, Z and Prosen, T and Zar\'and, Gergely},
  journal = {Phys. Rev. Lett.},
  volume = {135},
  issue = {10},
  pages = {100401},
  numpages = {7},
  year = {2025},
  month = {Sep},
  publisher = {American Physical Society},
  doi = {10.1103/f3c4-n21z},
  url = {https://link.aps.org/doi/10.1103/f3c4-n21z}
}

@article{devendra_2026,
  title = {Superdiffusion and anomalous fluctuations in chiral integrable dynamics},
  author = {Muzzi, Cristiano and Bhakuni, Devendra Singh and Dalmonte, Marcello and Zadnik, Lenart and Xavier, Hernan B.},
  journal = {Phys. Rev. B},
  volume = {113},
  issue = {18},
  pages = {184305},
  numpages = {12},
  year = {2026},
  month = {May},
  publisher = {American Physical Society},
  doi = {10.1103/q549-whw6},
  url = {https://link.aps.org/doi/10.1103/q549-whw6}
}

@article{Prosen_FV_2026,
  title = {Dynamic scaling and Family-Vicsek universality in $\mathrm{SU}(N)$ quantum spin chains},
  author = {Moca, C. P. and D\'ora, Bal\'azs and Sticlet, Doru and Valli, Angelo and Prosen, T. and Zar\'and, Gergely},
  journal = {Phys. Rev. B},
  volume = {113},
  issue = {2},
  pages = {L020405},
  numpages = {6},
  year = {2026},
  month = {Jan},
  publisher = {American Physical Society},
  doi = {10.1103/434y-71cd},
  url = {https://link.aps.org/doi/10.1103/434y-71cd}
}

@article{USJan2011,
title = {The density-matrix renormalization group in the age of matrix product states},
journal = {Annals of Physics},
volume = {326},
number = {1},
year = {2011},
url = {https://www.sciencedirect.com/science/article/pii/S0003491610001752},
author = {Ulrich Schollwöck},
}

@article{Landi_FCS_traj_2024,
  title = {Current Fluctuations in Open Quantum Systems: Bridging the Gap Between Quantum Continuous Measurements and Full Counting Statistics},
  author = {Landi, Gabriel T. and Kewming, Michael J. and Mitchison, Mark T. and Potts, Patrick P.},
  journal = {PRX Quantum},
  volume = {5},
  issue = {2},
  pages = {020201},
  numpages = {86},
  year = {2024},
  month = {Apr},
  publisher = {American Physical Society},
  doi = {10.1103/PRXQuantum.5.020201},
  url = {https://link.aps.org/doi/10.1103/PRXQuantum.5.020201}
}

@article{AMJFeb1982,
  title = {Nondiffusive Quantum Transport in a Dynamically Disordered Medium},
  author = {Jayannavar, A. M. and Kumar, N.},
  journal = {Phys. Rev. Lett.},
  volume = {48},
  issue = {8},
  pages = {553--556},
  numpages = {0},
  year = {1982},
  month = {Feb},
  publisher = {American Physical Society},
  doi = {10.1103/PhysRevLett.48.553},
  url = {https://link.aps.org/doi/10.1103/PhysRevLett.48.553}
}

@article{MBPJan1998,
  title = {The quantum-jump approach to dissipative dynamics in quantum optics},
  author = {Plenio, M. B. and Knight, P. L.},
  journal = {Rev. Mod. Phys.},
  volume = {70},
  issue = {1},
  pages = {101--144},
  numpages = {0},
  year = {1998},
  month = {Jan},
  publisher = {American Physical Society},
  doi = {10.1103/RevModPhys.70.101},
  url = {https://link.aps.org/doi/10.1103/RevModPhys.70.101}
}

@article{TPSep2012,
  title = {Diffusive high-temperature transport in the one-dimensional Hubbard model},
  author = {Prosen, Toma{\v{z}} and \ifmmode \check{Z}\else \v{Z}\fi{}nidari\ifmmode \check{c}\else \v{c}\fi{}, Marko},
  journal = {Phys. Rev. B},
  volume = {86},
  issue = {12},
  pages = {125118},
  numpages = {6},
  year = {2012},
  month = {Sep},
  publisher = {American Physical Society},
  doi = {10.1103/PhysRevB.86.125118},
  url = {https://link.aps.org/doi/10.1103/PhysRevB.86.125118}
}

@article{IROct2015,
doi = {10.1088/0034-4885/78/11/114001},
url = {https://dx.doi.org/10.1088/0034-4885/78/11/114001},
year = {2015},
month = {oct},
publisher = {IOP Publishing},
volume = {78},
number = {11},
pages = {114001},
author = {Rotter, I and Bird, J P},
title = {A review of progress in the physics of open quantum systems: theory and experiment},
journal = {Reports on Progress in Physics}
}

@article{ACApr2017,
  title = {Quantum Simulation of Generic Many-Body Open System Dynamics Using Classical Noise},
  author = {Chenu, A. and Beau, M. and Cao, J. and del Campo, A.},
  journal = {Phys. Rev. Lett.},
  volume = {118},
  issue = {14},
  pages = {140403},
  numpages = {6},
  year = {2017},
  month = {Apr},
  publisher = {American Physical Society},
  doi = {10.1103/PhysRevLett.118.140403},
  url = {https://link.aps.org/doi/10.1103/PhysRevLett.118.140403}
}

@article{SGJul2017,
  title = {Noise-Induced Subdiffusion in Strongly Localized Quantum Systems},
  author = {Gopalakrishnan, Sarang and Islam, K. Ranjibul and Knap, Michael},
  journal = {Phys. Rev. Lett.},
  volume = {119},
  issue = {4},
  pages = {046601},
  numpages = {6},
  year = {2017},
  month = {Jul},
  publisher = {American Physical Society},
  doi = {10.1103/PhysRevLett.119.046601},
  url = {https://link.aps.org/doi/10.1103/PhysRevLett.119.046601}
}

@article{APDec2021,
  title = {Subdiffusive Phases in Open Clean Long-Range Systems},
  author = {Purkayastha, Archak and Saha, Madhumita and Agarwalla, Bijay Kumar},
  journal = {Phys. Rev. Lett.},
  volume = {127},
  issue = {24},
  pages = {240601},
  numpages = {7},
  year = {2021},
  month = {Dec},
  publisher = {American Physical Society},
  doi = {10.1103/PhysRevLett.127.240601},
  url = {https://link.aps.org/doi/10.1103/PhysRevLett.127.240601}
}

@article{TGLDec2022,
  title = {Nonequilibrium boundary-driven quantum systems: Models, methods, and properties},
  author = {Landi, Gabriel T. and Poletti, Dario and Schaller, Gernot},
  journal = {Rev. Mod. Phys.},
  volume = {94},
  issue = {4},
  pages = {045006},
  numpages = {58},
  year = {2022},
  month = {Dec},
  publisher = {American Physical Society},
  doi = {10.1103/RevModPhys.94.045006},
  url = {https://link.aps.org/doi/10.1103/RevModPhys.94.045006}
}

@article{MSMay2023,
  title = {Universal Subdiffusive Behavior at Band Edges from Transfer Matrix Exceptional Points},
  author = {Saha, Madhumita and Agarwalla, Bijay Kumar and Kulkarni, Manas and Purkayastha, Archak},
  journal = {Phys. Rev. Lett.},
  volume = {130},
  issue = {18},
  pages = {187101},
  numpages = {6},
  year = {2023},
  month = {May},
  publisher = {American Physical Society},
  doi = {10.1103/PhysRevLett.130.187101},
  url = {https://link.aps.org/doi/10.1103/PhysRevLett.130.187101}
}

@article{MSOct2023,
  title = {Environment assisted superballistic scaling of conductance},
  author = {Saha, Madhumita and Agarwalla, Bijay Kumar and Kulkarni, Manas and Purkayastha, Archak},
  journal = {Phys. Rev. B},
  volume = {108},
  issue = {16},
  pages = {L161115},
  numpages = {6},
  year = {2023},
  month = {Oct},
  publisher = {American Physical Society},
  doi = {10.1103/PhysRevB.108.L161115},
  url = {https://link.aps.org/doi/10.1103/PhysRevB.108.L161115}
}

@article{MAP2025,
      title={Quantum Dynamics with Stochastic Non-Hermitian Hamiltonians}, 
      author={Pablo Martinez-Azcona and Aritra Kundu and Avadh Saxena and Adolfo del Campo and Aurelia Chenu},
      year={2025},
      journal = {arXiv:2407.07746},
      url={https://arxiv.org/abs/2407.07746}, 
}

@article{PN2025,
    title = {Quantum dynamics in Krylov space: Methods and applications},
    journal = {Physics Reports},
    volume = {1125-1128},
    year = {2025},
    url = {https://www.sciencedirect.com/science/article/pii/S0370157325001462},
    author = {Pratik Nandy and Apollonas S. Matsoukas-Roubeas and Pablo Martínez-Azcona and Anatoly Dymarsky and Adolfo {del Campo}},
}

@article{RS2025,
  title={Dephasing enabled fast charging of quantum batteries},
  author={Shastri, Rahul and Jiang, Chao and Xu, Guo-Hua and Prasanna Venkatesh, B and Watanabe, Gentaro},
  journal={npj Quantum Information},
  volume={11},
  number={1},
  pages={9},
  year={2025},
  url={https://doi.org/10.1038/s41534-025-00959-5},
  publisher={Nature Publishing Group UK London}
}

@article{AH2012,
  title={On-chip quantum simulation with superconducting circuits},
  author={Houck, Andrew A and T{\"u}reci, Hakan E and Koch, Jens},
  journal={Nature Physics},
  volume={8},
  number={4},
  pages={292--299},
  year={2012},
  publisher={Nature Publishing Group UK London},
  url={https://doi.org/10.1038/nphys2251}
}

@article{IB2012,
  title={Quantum simulations with ultracold quantum gases},
  author={Bloch, Immanuel and Dalibard, Jean and Nascimbene, Sylvain},
  journal={Nature Physics},
  volume={8},
  number={4},
  pages={267--276},
  year={2012},
  publisher={Nature Publishing Group},
  url={https://doi.org/10.1038/nphys2259}
}

@article{RB2012,
  title={Quantum simulations with trapped ions},
  author={Blatt, Rainer and Roos, Christian F},
  journal={Nature Physics},
  volume={8},
  number={4},
  pages={277--284},
  year={2012},
  publisher={Nature Publishing Group UK London},
  url={https://doi.org/10.1038/nphys2252}
}

@article{CM2019,
  title = {Environment-Assisted Quantum Transport in a 10-qubit Network},
  author = {Maier, Christine and Brydges, Tiff and Jurcevic, Petar and Trautmann, Nils and Hempel, Cornelius and Lanyon, Ben P. and Hauke, Philipp and Blatt, Rainer and Roos, Christian F.},
  journal = {Phys. Rev. Lett.},
  volume = {122},
  issue = {5},
  pages = {050501},
  numpages = {6},
  year = {2019},
  month = {Feb},
  publisher = {American Physical Society},
  doi = {10.1103/PhysRevLett.122.050501},
  url = {https://link.aps.org/doi/10.1103/PhysRevLett.122.050501}
}

@article{XZJan2025,
  title={Learning and forecasting open quantum dynamics with correlated noise},
  author={Zhang, Xinfang and Wu, Zhihao and White, Gregory AL and Xiang, Zhongcheng and Hu, Shun and Peng, Zhihui and Liu, Yong and Zheng, Dongning and Fu, Xiang and Huang, Anqi and Poletti, Dario and Modi, Kavan and Wu, Junjie and Deng, Mingtang and Guo, Che},
  journal={Communications Physics},
  volume={8},
  number={1},
  pages={29},
  year={2025},
  publisher={Nature Publishing Group UK London},
  url={https://doi.org/10.1038/s42005-025-01944-2}
}

@article{CFRoos_2025,
  title = {Measuring Full Counting Statistics in a Trapped-Ion Quantum Simulator},
  author = {Joshi, Lata Kh and Ares, Filiberto and Joshi, Manoj K. and Roos, Christian F. and Calabrese, Pasquale},
  journal = {Phys. Rev. Lett.},
  volume = {135},
  issue = {16},
  pages = {160601},
  numpages = {6},
  year = {2025},
  month = {Oct},
  publisher = {American Physical Society},
  doi = {10.1103/gyvf-s5bd},
  url = {https://link.aps.org/doi/10.1103/gyvf-s5bd}
}

@article{MK2022,
author = {M. K. Joshi  and F. Kranzl  and A. Schuckert  and I. Lovas  and C. Maier  and R. Blatt  and M. Knap  and C. F. Roos },
title = {Observing emergent hydrodynamics in a long-range quantum magnet},
journal = {Science},
volume = {376},
number = {6594},
pages = {720-724},
year = {2022},
doi = {10.1126/science.abk2400},
URL = {https://www.science.org/doi/abs/10.1126/science.abk2400}}

@article{Sarang2025,
  title = {Superdiffusive Transport in Chaotic Quantum Systems with Nodal Interactions},
  author = {Wang, Yu-Peng and Ren, Jie and Gopalakrishnan, Sarang and Vasseur, Romain},
  journal = {Phys. Rev. Lett.},
  volume = {135},
  issue = {16},
  pages = {166303},
  numpages = {6},
  year = {2025},
  month = {Oct},
  publisher = {American Physical Society},
  doi = {10.1103/xx9z-4j6c},
  url = {https://link.aps.org/doi/10.1103/xx9z-4j6c}
}

@article{Immanuel2024,
	author = {Wienand, Julian F. and Karch, Simon and Impertro, Alexander and Schweizer, Christian and McCulloch, Ewan and Vasseur, Romain and Gopalakrishnan, Sarang and Aidelsburger, Monika and Bloch, Immanuel},
	journal = {Nature Physics},
	number = {11},
	title = {Emergence of fluctuating hydrodynamics in chaotic quantum systems},
	url = {https://doi.org/10.1038/s41567-024-02611-z},
	volume = {20},
	year = {2024}}

@article{L1976,
  title={On the generators of quantum dynamical semigroups},
  author={Lindblad, Goran},
  journal={Communications in mathematical physics},
  volume={48},
  number={2},
  pages={119--130},
  year={1976},
  publisher={Springer},
  url = {https://projecteuclid.org/journals/communications-in-mathematical-physics/volume-48/issue-2/On-the-generators-of-quantum-dynamical-semigroups/cmp/1103899849.full}
}

@article{VG1976,
  title={Completely positive dynamical semigroups of N-level systems},
  author={Gorini, Vittorio and Kossakowski, Andrzej and Sudarshan, Ennackal Chandy George},
  journal={Journal of Mathematical Physics},
  volume={17},
  number={5},
  pages={821--825},
  year={1976},
  publisher={American Institute of Physics},
  url={https://doi.org/10.1063/1.522979}
}

@article{Rainer1986,
  title = {Observation of Quantum Jumps},
  author = {Sauter, Th. and Neuhauser, W. and Blatt, R. and Toschek, P. E.},
  journal = {Phys. Rev. Lett.},
  volume = {57},
  issue = {14},
  pages = {1696--1698},
  year = {1986},
  month = {Oct},
  publisher = {American Physical Society},
  doi = {10.1103/PhysRevLett.57.1696},
  url = {https://link.aps.org/doi/10.1103/PhysRevLett.57.1696}
}

@article{Vijay2011,
  title = {Observation of Quantum Jumps in a Superconducting Artificial Atom},
  author = {Vijay, R. and Slichter, D. H. and Siddiqi, I.},
  journal = {Phys. Rev. Lett.},
  volume = {106},
  issue = {11},
  pages = {110502},
  numpages = {4},
  year = {2011},
  month = {Mar},
  publisher = {American Physical Society},
  doi = {10.1103/PhysRevLett.106.110502},
  url = {https://link.aps.org/doi/10.1103/PhysRevLett.106.110502}
}

@article{Kater2013,
	author = {Murch, K. W. and Weber, S. J. and Macklin, C. and Siddiqi, I.},
	journal = {Nature},
	number = {7470},
	title = {Observing single quantum trajectories of a superconducting quantum bit},
	url = {https://doi.org/10.1038/nature12539},
	volume = {502},
	year = {2013}}

@article{dutta2025,
      title={An introduction to Markovian open quantum systems}, 
      author={Shovan Dutta},
      year={2025},
      journal = {arXiv:2510.26530},
      url={https://arxiv.org/abs/2510.26530}, 
}

@book{BPOQS,
    author = {Breuer, Heinz-Peter and Petruccione, Francesco},
    title = {The Theory of Open Quantum Systems},
    publisher = {Oxford University Press},
    year = {2007},
    month = {01},
    isbn = {9780199213900},
    doi = {10.1093/acprof:oso/9780199213900.001.0001},
    url = {https://doi.org/10.1093/acprof:oso/9780199213900.001.0001},
}

@book{carmichael2009,
  title={An Open Systems Approach to Quantum Optics: Lectures Presented at the Universit{\'e} Libre de Bruxelles, October 28 to November 4, 1991},
  author={Carmichael, H.},
  url={https://books.google.co.in/books?id=uor_CAAAQBAJ},
  year={2009},
  publisher={Springer Berlin Heidelberg}
}

@article{ToddBrun2000,
  title = {Continuous measurements, quantum trajectories, and decoherent histories},
  author = {Brun, Todd A.},
  journal = {Phys. Rev. A},
  volume = {61},
  issue = {4},
  pages = {042107},
  numpages = {17},
  year = {2000},
  month = {Mar},
  publisher = {American Physical Society},
  doi = {10.1103/PhysRevA.61.042107},
  url = {https://link.aps.org/doi/10.1103/PhysRevA.61.042107}
}

@book{wiseman2010,
  title={Quantum Measurement and Control},
  author={Wiseman, H.M. and Milburn, G.J.},
  isbn={9780521804424},
  lccn={2009034266},
  url={https://books.google.co.in/books?id=ZNjvHaH8qA4C},
  year={2010},
  publisher={Cambridge University Press}
}

@book{jacobs2014,
  title={Quantum Measurement Theory and its Applications},
  author={Jacobs, K.},
  isbn={9781107025486},
  lccn={2014011297},
  url={https://books.google.co.in/books?id=gzTRngEACAAJ},
  year={2014},
  publisher={Cambridge University Press}
}

@article{Jacobs01092006,
author = {Kurt Jacobs and Daniel A. Steck},
title = {A straightforward introduction to continuous quantum measurement},
journal = {Contemporary Physics},
volume = {47},
number = {5},
pages = {279--303},
year = {2006},
publisher = {Taylor \& Francis},
doi = {10.1080/00107510601101934},
URL = { https://doi.org/10.1080/00107510601101934
}}

@article{NGisin_1992,
doi = {10.1088/0305-4470/25/21/023},
url = {https://doi.org/10.1088/0305-4470/25/21/023},
year = {1992},
month = {nov},
publisher = {},
volume = {25},
number = {21},
pages = {5677},
author = {N Gisin and I C Percival},
title = {The quantum-state diffusion model applied to open systems},
journal = {Journal of Physics A: Mathematical and General}}

@article{Yuen_83,
author = {Horace P. Yuen and Vincent W. S. Chan},
journal = {Opt. Lett.},
number = {3},
pages = {177--179},
title = {Noise in homodyne and heterodyne detection},
volume = {8},
year = {1983},
url = {https://opg.optica.org/ol/abstract.cfm?URI=ol-8-3-177},
doi = {10.1364/OL.8.000177}}

@article{Warszawski_2002,
   title={Quantum trajectories for realistic photodetection: I. General formalism},
   volume={5},
   url={http://dx.doi.org/10.1088/1464-4266/5/1/301},
   number={1},
   journal={Journal of Optics B: Quantum and Semiclassical Optics},
   author={Warszawski, P and Wiseman, H M},
   year={2002}}

@article{Molmer1992,
  title = {Wave-function approach to dissipative processes in quantum optics},
  author = {Dalibard, Jean and Castin, Yvan and M\o{}lmer, Klaus},
  journal = {Phys. Rev. Lett.},
  volume = {68},
  issue = {5},
  pages = {580--583},
  numpages = {0},
  year = {1992},
  month = {Feb},
  publisher = {American Physical Society},
  doi = {10.1103/PhysRevLett.68.580},
  url = {https://link.aps.org/doi/10.1103/PhysRevLett.68.580}
}

@article{GISIN1992315,
title = {Wave-function approach to dissipative processes: are there quantum jumps?},
journal = {Physics Letters A},
volume = {167},
number = {4},
pages = {315-318},
year = {1992},
issn = {0375-9601},
doi = {https://doi.org/10.1016/0375-9601(92)90264-M},
url = {https://www.sciencedirect.com/science/article/pii/037596019290264M},
author = {Nicolas Gisin and Ian C. Percival},
}

@article{Bijay2012,
  title = {Full-counting statistics of heat transport in harmonic junctions: Transient, steady states, and fluctuation theorems},
  author = {Agarwalla, Bijay Kumar and Li, Baowen and Wang, Jian-Sheng},
  journal = {Phys. Rev. E},
  volume = {85},
  issue = {5},
  pages = {051142},
  numpages = {19},
  year = {2012},
  month = {May},
  publisher = {American Physical Society},
  doi = {10.1103/PhysRevE.85.051142},
  url = {https://link.aps.org/doi/10.1103/PhysRevE.85.051142}
}

@article{ganguly2026,
      title={Measurement induced faster symmetry restoration in quantum trajectories}, 
      author={Katha Ganguly and Bijay Kumar Agarwalla},
      year={2026},
      journal={arXiv:2601.18458},
      url={https://arxiv.org/abs/2601.18458}
}

@Article{Xhek2025,
AUTHOR = {Di Giulio, Giuseppe and Turkeshi, Xhek and Murciano, Sara},
TITLE = {Measurement-Induced Symmetry Restoration and Quantum Mpemba Effect},
JOURNAL = {Entropy},
VOLUME = {27},
YEAR = {2025},
NUMBER = {4},
ARTICLE-NUMBER = {407},
URL = {https://www.mdpi.com/1099-4300/27/4/407}
}

@article{Sarang_2022,
  title = {Entanglement and Charge-Sharpening Transitions in U(1) Symmetric Monitored Quantum Circuits},
  author = {Agrawal, Utkarsh and Zabalo, Aidan and Chen, Kun and Wilson, Justin H. and Potter, Andrew C. and Pixley, J. H. and Gopalakrishnan, Sarang and Vasseur, Romain},
  journal = {Phys. Rev. X},
  volume = {12},
  issue = {4},
  pages = {041002},
  numpages = {29},
  year = {2022},
  month = {Oct},
  publisher = {American Physical Society},
  doi = {10.1103/PhysRevX.12.041002},
  url = {https://link.aps.org/doi/10.1103/PhysRevX.12.041002}
}

@article{Feng_2025,
   title={Charge and Spin Sharpening Transitions on Dynamical Quantum Trees},
   volume={9},
   url={http://dx.doi.org/10.22331/q-2025-04-07-1692},
   DOI={10.22331/q-2025-04-07-1692},
   journal={Quantum},
   author={Feng, Xiaozhou and Fishchenko, Nadezhda and Gopalakrishnan, Sarang and Ippoliti, Matteo},
   year={2025},
   month=Apr, pages={1692} }

@article{Skinner2019,
  title = {Measurement-Induced Phase Transitions in the Dynamics of Entanglement},
  author = {Skinner, Brian and Ruhman, Jonathan and Nahum, Adam},
  journal = {Phys. Rev. X},
  volume = {9},
  issue = {3},
  pages = {031009},
  numpages = {21},
  year = {2019},
  month = {Jul},
  publisher = {American Physical Society},
  doi = {10.1103/PhysRevX.9.031009},
  url = {https://link.aps.org/doi/10.1103/PhysRevX.9.031009}
}

@article{sebastian_2021,
  title = {Entanglement Transition in a Monitored Free-Fermion Chain: From Extended Criticality to Area Law},
  author = {Alberton, O. and Buchhold, M. and Diehl, S.},
  journal = {Phys. Rev. Lett.},
  volume = {126},
  issue = {17},
  pages = {170602},
  numpages = {6},
  year = {2021},
  url = {https://link.aps.org/doi/10.1103/PhysRevLett.126.170602}
}

@article{Ashida2020,
  title = {Measurement-induced quantum criticality under continuous monitoring},
  author = {Fuji, Yohei and Ashida, Yuto},
  journal = {Phys. Rev. B},
  volume = {102},
  issue = {5},
  pages = {054302},
  numpages = {14},
  year = {2020},
  month = {Aug},
  publisher = {American Physical Society},
  doi = {10.1103/PhysRevB.102.054302},
  url = {https://link.aps.org/doi/10.1103/PhysRevB.102.054302}
}

@article{GauthameshwarPoletti2025,
  title = {Quantum thermal machines and the emergence of different thermodynamic functioning regimes from finite coupling to a load},
  author = {S., Gauthameshwar and Jaseem, Noufal and Poletti, Dario},
  journal = {Phys. Rev. A},
  volume = {112},
  issue = {5},
  pages = {L050201},
  numpages = {6},
  year = {2025},
  month = {Nov},
  publisher = {American Physical Society},
  doi = {10.1103/dr9b-5ryh},
  url = {https://link.aps.org/doi/10.1103/dr9b-5ryh}
}

@book{Binder2018,
  title     = {Thermodynamics in the Quantum Regime: Fundamental Aspects and New Directions},
  editor    = {Binder, Felix and Correa, Luis A. and Gogolin, Christian and Anders, Janet and Adesso, Gerardo},
  series    = {Fundamental Theories of Physics},
  volume    = {195},
  year      = {2018},
  publisher = {Springer International Publishing},
  doi       = {10.1007/978-3-319-99046-0},
  isbn      = {978-3-319-99045-3}
}

@article{Kessler_2012_Dissipative,
  title     = {Dissipative phase transitions},
  author    = {Kessler, E. M. and Giedke, G. and Imamoglu, A. and Yelin, S. F. and Lukin, M. D. and Cirac, J. I.},
  journal   = {Phys. Rev. A},
  volume    = {86},
  issue     = {1},
  pages     = {012116},
  year      = {2012},
  publisher = {American Physical Society},
  doi       = {10.1103/PhysRevA.86.012116}
}

@article{Fitzpatrick_2017_OpenExp,
  title     = {Observation of a Dissipative Phase Transition in a Many-Body Cavity QED System},
  author    = {Fitzpatrick, M. and Sundaresan, N. M. and Li, A. C. Y. and Koch, J. and Houck, A. A.},
  journal   = {Phys. Rev. X},
  volume    = {7},
  issue     = {1},
  pages     = {011016},
  year      = {2017},
  publisher = {American Physical Society},
  doi       = {10.1103/PhysRevX.7.011016}
}

@article{Daley_2014_Review,
  title     = {Quantum trajectories and open many-body quantum systems},
  author    = {Daley, Andrew J.},
  journal   = {Advances in Physics},
  volume    = {63},
  number    = {2},
  pages     = {77--149},
  year      = {2014},
  publisher = {Taylor \& Francis},
  doi       = {10.1080/00018732.2014.933502},
  url       = {https://arxiv.org/abs/1405.6694}
}

@article{Lange_2024_NQS_Review,
  title     = {From architectures to applications: a review of neural quantum states},
  author    = {Lange, Hannah and Van de Walle, Anka and Abedinnia, Atiye and Bohrdt, Annabelle},
  journal   = {Quantum Science and Technology},
  volume    = {9},
  number    = {4},
  pages     = {040501},
  year      = {2024},
  month     = {Sep},
  publisher = {IOP Publishing},
  doi       = {10.1088/2058-9565/ad7168},
  url       = {https://iopscience.iop.org/article/10.1088/2058-9565/ad7168}
}

@article{Hartmann_Carleo_2019,
  title     = {Neural-Network Approach to Dissipative Quantum Many-Body Dynamics},
  author    = {Hartmann, Michael J. and Carleo, Giuseppe},
  journal   = {Phys. Rev. Lett.},
  volume    = {122},
  number    = {25},
  pages     = {250502},
  year      = {2019},
  month     = {Jun},
  publisher = {American Physical Society},
  doi       = {10.1103/PhysRevLett.122.250502},
  url       = {https://link.aps.org/doi/10.1103/PhysRevLett.122.250502}
}

@article{Vicentini_2019_Purified,
  title     = {Variational Neural-Network Ansatz for Steady States in Open Quantum Systems},
  author    = {Vicentini, Filippo and Biella, Alberto and Regnault, Nicolas and Ciuti, Cristiano},
  journal   = {Phys. Rev. Lett.},
  volume    = {122},
  number    = {25},
  pages     = {250503},
  year      = {2019},
  month     = {Jun},
  publisher = {American Physical Society},
  doi       = {10.1103/PhysRevLett.122.250503},
  url       = {https://link.aps.org/doi/10.1103/PhysRevLett.122.250503}
}

@article{Yoshioka_Hamazaki_2019,
  title     = {Constructing neural stationary states for open quantum many-body systems},
  author    = {Yoshioka, Nobuyuki and Hamazaki, Ryusuke},
  journal   = {Phys. Rev. B},
  volume    = {99},
  number    = {21},
  pages     = {214306},
  year      = {2019},
  month     = {Jun},
  publisher = {American Physical Society},
  doi       = {10.1103/PhysRevB.99.214306},
  url       = {https://link.aps.org/doi/10.1103/PhysRevB.99.214306}
}

@article{Luo_2022_Autoregressive,
  title     = {Autoregressive Neural Network for Simulating Open Quantum System Dynamics},
  author    = {Luo, Di and Zhao, Jiayu and Clark, Bryan K.},
  journal   = {Phys. Rev. Lett.},
  volume    = {128},
  number    = {9},
  pages     = {090501},
  year      = {2022},
  month     = {Feb},
  publisher = {American Physical Society},
  doi       = {10.1103/PhysRevLett.128.090501},
  url       = {https://link.aps.org/doi/10.1103/PhysRevLett.128.090501}
}

@article{Reh_2021_AutoregressiveOpen,
  title     = {Time-dependent variational principle for open quantum systems with artificial neural networks},
  author    = {Reh, Moritz and Schmitt, Markus and G{\"a}rttner, Martin},
  journal   = {Phys. Rev. Lett.},
  volume    = {127},
  number    = {23},
  pages     = {230501},
  year      = {2021},
  month     = {Nov},
  publisher = {American Physical Society},
  doi       = {10.1103/PhysRevLett.127.230501},
  url       = {https://aps.org}
}

@article{Vicentini_2022_GHDO,
  title     = {Positive-definite parametrization of mixed quantum states with deep neural networks},
  author    = {Vicentini, Filippo and Rossi, Riccardo and Carleo, Giuseppe},
  journal   = {SciPost Physics},
  volume    = {12},
  number    = {3},
  pages     = {100},
  year      = {2022},
  publisher = {SciPost Foundation},
  doi       = {10.21468/SciPostPhys.12.3.100},
  url       = {https://arxiv.org/abs/2206.13488}
}

@article{Zhang_2025_NQP,
  title     = {Neural quantum propagators for driven-dissipative quantum dynamics},
  author    = {Zhang, Jiaji and Benavides-Riveros, Carlos L. and Chen, Lipeng},
  journal   = {Phys. Rev. Res},
  volume    = {7},
  number    = {1},
  pages     = {L012013},
  year      = {2025},
  month     = {Jan},
  publisher = {American Physical Society},
  doi       = {10.1103/PhysRevResearch.7.L012013},
  url       = {https://link.aps.org/doi/10.1103/PhysRevResearch.7.L012013}
}

@article{Sinibaldi_2026_NeuralGalerkin,
  title     = {Time-Dependent Neural Galerkin Method for Quantum Dynamics},
  author    = {Sinibaldi, Alessandro and Hendry, Douglas and Vicentini, Filippo and Carleo, Giuseppe},
  journal   = {Phys. Rev. Lett.},
  volume    = {136},
  number    = {12},
  pages     = {120402},
  year      = {2026},
  month     = {March},
  publisher = {American Physical Society},
  doi = {10.1103/kqvx-dl54},
  url = {https://link.aps.org/doi/10.1103/kqvx-dl54}
}

@article{Schmitt_2025_tNQS,
  title     = {Many-body dynamics with explicitly time-dependent neural quantum states},
  author    = {Van de Walle, Anka and Schmitt, Markus and Bohrdt, Annabelle},
  journal   = {Machine Learning: Science and Technology},
  volume    = {6},
  number    = {4},
  pages     = {045011},
  year      = {2025},
  month     = {October},
  publisher = {IOP Publishing},
  doi       = {10.1088/2632-2153/ae0f39},
  url       = {https://iopscience.iop.org/article/10.1088/2632-2153/ae0f39}
}

@article{Hou_2026_SpacetimeTDSE,
  title     = {A global spacetime optimization approach to the real-space time-dependent Sch{\"o}dinger equation},
  author    = {Hou, Enze and Liu, Yuzhi and Zhang, Linxuan and Ye, Difa and Wang, Lei and Wang, Han},
  journal   = {Machine Learning: Science and Technology},
  volume    = {7},
  number    = {3},
  pages     = {035012},
  year      = {2026},
  publisher = {IOP Publishing},
  doi       = {10.1088/2632-2153/ae62c8},
  url       = {https://iopscience.iop.org/article/10.1088/2632-2153/ae62c8}
}

@article{wang2026continuoustimeparametrizationneuralquantum,
      title={Continuous-time parametrization of neural quantum states for quantum dynamics}, 
      author={Dingzu Wang and Wenxuan Zhang and Xiansong Xu and Dario Poletti},
      year={2026},
      journal = {arXiv:2507.08418},
      url={https://arxiv.org/abs/2507.08418}, 
}

@article{Jiang_2026,
doi = {10.1088/0256-307X/43/7/070302},
url = {https://doi.org/10.1088/0256-307X/43/7/070302},
year = {2026},
month = {jul},
publisher = {Chinese Physical Society and IOP Publishing Ltd},
volume = {43},
number = {7},
pages = {070302},
author = {Jiang, Shixian and Liu, Jianpeng and Yuan, Jianmin and Guan, Xi-Wen and Li, Yongqiang},
title = {Universal Scaling of Higher-Order Cumulants in Quantum Isotropic Spin Chains},
journal = {Chinese Physics Letters}}

@article{Ray_2026,
  title = {Quantum dynamics in lattices in the presence of bulk dephasing and a localized source},
  author = {Ray, Tamoghna and Ganguly, Katha and Poletti, Dario and Kulkarni, Manas and Agarwalla, Bijay Kumar},
  journal = {Phys. Rev. B},
  volume = {113},
  issue = {5},
  pages = {054307},
  numpages = {11},
  year = {2026},
  doi = {10.1103/6frd-chqr},
  url = {https://link.aps.org/doi/10.1103/6frd-chqr}
}

@article{Ganguly2024,
  title = {Transport in open quantum systems in the presence of lossy channels},
  author = {Ganguly, Katha and Kulkarni, Manas and Agarwalla, Bijay Kumar},
  journal = {Phys. Rev. B},
  volume = {110},
  issue = {23},
  pages = {235425},
  numpages = {20},
  year = {2024},
  month = {Dec},
  publisher = {American Physical Society},
  doi = {10.1103/PhysRevB.110.235425},
  url = {https://link.aps.org/doi/10.1103/PhysRevB.110.235425}
}

@article{Fujimoto2022,
  title = {Impact of Dissipation on Universal Fluctuation Dynamics in Open Quantum Systems},
  author = {Fujimoto, Kazuya and Hamazaki, Ryusuke and Kawaguchi, Yuki},
  journal = {Phys. Rev. Lett.},
  volume = {129},
  issue = {11},
  pages = {110403},
  numpages = {7},
  year = {2022},
  month = {Sep},
  publisher = {American Physical Society},
  doi = {10.1103/PhysRevLett.129.110403},
  url = {https://link.aps.org/doi/10.1103/PhysRevLett.129.110403}
}

@article{Archak_2018,
  title = {Anomalous transport in the Aubry-Andr\'e-Harper model in isolated and open systems},
  author = {Purkayastha, Archak and Sanyal, Sambuddha and Dhar, Abhishek and Kulkarni, Manas},
  journal = {Phys. Rev. B},
  volume = {97},
  issue = {17},
  pages = {174206},
  numpages = {11},
  year = {2018},
  month = {May},
  publisher = {American Physical Society},
  doi = {10.1103/PhysRevB.97.174206},
  url = {https://link.aps.org/doi/10.1103/PhysRevB.97.174206}
}

@article{Fazio2025,
	title = {Many-body open quantum systems},
	pages = {99},
	author = {Fazio, Rosario and Keeling, Jonathan and Mazza, Leonardo and Schirò, Marco},
	journal = {SciPost Phys. Lect. Notes},
	year = {2025},
	publisher = {SciPost},
	doi = {10.21468/SciPostPhysLectNotes.99},
	url = {https://scipost.org/10.21468/SciPostPhysLectNotes.99}
}

@article{supp_mat,
  title = {},
  author = {},
  journal = {See Supplementary Material},
  volume = {},
  issue = {},
  pages = {},
  numpages = {},
  year = {},
  month = {},
  publisher = {},
  doi = {},
  url = {}
}

\newpage
\onecolumngrid

\setcounter{figure}{0}
\renewcommand{\thefigure}{S\arabic{figure}}
\setcounter{equation}{0}
\renewcommand{\theequation}{S\arabic{equation}}

\begin{center}
\textbf{Supplemental Material for ``Long-time Dynamics of Many-body Open Quantum Systems using Quantum Generating Functions"}
\end{center}

\section{Proof of the equivalence between the full counting statistics of the ensemble-averaged dynamics and the distribution obtained from quantum jumps}
In this section, we prove the equivalence between the FCS obtained from the ensemble-averaged GKSL dynamics and the probability distribution obtained from quantum jump trajectories, i.e., $P(n,t)=P_{\rm QJ}(n,t)$, for setup-I. Recall that the GKSL equation governing the ensemble-averaged dynamics of the open quantum many-body system under local particle injection is given by
\begin{align}
\frac{d\rho}{dt}=\mathcal{L}\rho=-i[\hat{H},\rho]+\Gamma_G\Big(\hat{S}_1^+\rho \hat{S}_1^--\frac{1}{2}\{\hat{S}_1^-\hat{S}_1^+,\rho\}\Big). \label{eq:qgf_supp}
\end{align}
For this dynamics, the quantum generating function (QGF) is defined as $G(\lambda,t)=\mathrm{Tr}\big[e^{i\lambda \hat{S}^z}e^{\mathcal{L}t}[e^{-i\lambda \hat{S}^z}\rho_0]\big]$. Since the GKSL dynamics can be unraveled into stochastic quantum trajectories, we write $e^{\mathcal{L}t}[\rho]=\sum_{\xi}\hat{M}_{\xi}\rho \hat{M}_{\xi}^{\dagger}$, where $\xi$ labels a stochastic trajectory. The QGF can therefore be expressed as
\begin{align}
G(\lambda,t)=\sum_{\xi}\mathrm{Tr}\Big[e^{i\lambda \hat{S}^z}\hat{M}_{\xi}[e^{-i\lambda \hat{S}^z}\rho_0]\hat{M}_{\xi}^{\dagger}\Big].
\end{align}
A stochastic trajectory can be represented by a sequence of stochastic outcomes ${\mu_1,\mu_2,\dots,\mu_t}$ corresponding to each time step $dt$ between $0$ and $t$. The variables $\mu_i$ depend on the particular unraveling protocol. For example, in a continuous measurement protocol, they correspond to the measurement outcomes which are either $0$ (no jump event) or $1$ (jump). For the quantum jump unraveling, the trajectory operator is given by $\hat{M}_{\xi}=\hat{M}_{\mu_1}\hat{M}_{\mu_2}\dots \hat{M}_{\mu_t}$. Here, if $\mu_j$ corresponds to a no-jump event, then $\hat{M}_{\mu_j}=\mathbb{\hat{I}}-i\hat{H}dt-\frac{\Gamma_G dt}{2}\hat{S}_1^{-}\hat{S}_1^{+}$, whereas if $\mu_j$ corresponds to a jump event, then $\hat{M}_{\mu_j}=\sqrt{\Gamma_G dt}\hat{S}_1^{+}$. For a no-jump event,
$e^{i\lambda \hat{S}^z}\hat{M}_{\mu_j}e^{-i\lambda \hat{S}^z}=\hat{M}_{\mu_j}$, while for a jump event,
$e^{i\lambda \hat{S}^z}\hat{M}_{\mu_j}e^{-i\lambda \hat{S}^z}=e^{i\lambda}\hat{M}_{\mu_j}$. If the $\xi$-th trajectory contains $n_\xi$ number of jump events, then
$e^{i\lambda \hat{S}^z}\hat{M}_{\xi}e^{-i\lambda \hat{S}^z}=e^{i\lambda n_\xi}\hat{M}_{\xi}$.
Consequently, the QGF becomes
\begin{align}
G(\lambda,t)=\sum_\xi e^{i\lambda n_\xi} \, \mathrm{Tr}[\hat{M}_\xi\rho_0 \hat{M}_\xi^{\dagger}]=\sum_\xi e^{i\lambda n_\xi}\, P(\xi), \label{eq:G_lambda_supp_sim}
\end{align}
where $P(\xi)$ is the probability of the $\xi$-th quantum jump trajectory. 
Now performing the Fourier transform in Eq.~\eqref{eq:G_lambda_supp_sim} yields
\begin{align}
P(n,t)=\sum_\xi \delta_{n,n_\xi} P(\xi). \label{eq:GKSL_Pn_supp}
\end{align}
Eq.~\eqref{eq:GKSL_Pn_supp} gives the probability distribution of the total injected charge into the system for the dynamics governed by the GKSL equation. An alternative approach is to sample the injected charge directly from individual quantum jump trajectories.
Since each jump injects one excitation into the system, the total number of jumps in a trajectory is equal to the change in the total excitation number. Hence, we can write $n_\xi=\la \hat{S}^z\ra_t^{\xi}-\la\hat{S}^z\ra_0$. The probability distribution of the injected charge obtained from quantum jump trajectories is therefore
\begin{align}
P_{\rm QJ}(n,t)=\sum_\xi \delta_{n,\la {\hat{S}^z}\ra_t^{\xi}-\la{\hat{S}^z}\ra_0} P(\xi)=P(n,t). \label{eq:jump_pn_supp}
\end{align}
Thus, we conclude that $P_{\rm QJ}(n,t)=P(n,t)$.

It is important to note that this equivalence does not hold for quantum state diffusion (QSD) trajectories. If one samples the total injected charge from QSD trajectories to construct the distribution $P_{\rm QSD}(n,t)$, the resulting distribution is, in general, different from $P(n,t)$. The reason is that, unlike the quantum jump unraveling, the QSD measurement operators do not satisfy the relation $e^{i\lambda \hat{S}^z}\hat{M}_\mu e^{-i\lambda \hat{S}^z}=e^{i\lambda}\hat{M}_\mu$. Therefore, the counting field cannot be factored out in the same manner, and the above derivation no longer applies.
\section{Convergence of moments with counting field}
In this section, we discuss the convergence of the moments $\mu_n(t)$ with the counting field $\lambda$. Recall that the moments are obtained from the generating function $G(\lambda,t)$ by expanding it in powers of $\lambda$ as
\begin{align}
G(\lambda,t)=1+\sum_{n=1}^{\infty}\frac{(i\lambda)^n}{n!}\mu_n(t).
\label{eq:G_lamda_exp_supp}
\end{align}
As we are only interested in the first and second moments of the net transferred charge, i.e., $\mu_1(t)$ and $\mu_2(t)$, we retain the terms up to the lowest order in $\lambda$ to compute real and imaginary parts of $G(\lambda,t)$. The expressions for $\mu_1(t)$ and $\mu_2(t)$ are obtained from $G(\lambda,t)$ as
\begin{align}
&\mu_1(t)=\frac{1}{\lambda}\mathrm{Im}[G(\lambda,t)] + O(\lambda^2),
\label{eq:mu_1_supp}\\
&\mu_2(t)=\frac{2}{\lambda^2}\left(1-\mathrm{Re}[G(\lambda,t)]\right)+O(\lambda^2).
\label{eq:mu_2_supp}
\end{align}
Hence, an obvious choice for simulating the dynamics is to use small values of $\lambda$. Also for small $\lambda$, the counting operator $R(\lambda,t)$ remains very close to Identity ($\mathbb{I}$), thereby facilitating the simulation with less bond dimension. Another reason to choose small $\lambda$ is, for a given $\lambda$, the expansion in Eq.~\eqref{eq:G_lamda_exp_supp} is valid only up to a certain time $t$, within which
\begin{align}
\lambda\mu_1(t)\gg\lambda^3\mu_3(t),\qquad
\lambda^2\mu_2(t)\gg\lambda^4\mu_4(t).
\end{align}
Therefore, at very large times, when the moments $\mu_n(t)$ become larger in magnitude, the perturbative expansion in Eq.~\eqref{eq:G_lamda_exp_supp} eventually breaks down. Thus, to simulate the dynamics up to a sufficiently long time $t$, one needs to choose a sufficiently small $\lambda$.
\begin{figure}[h!]
    \centering
    \includegraphics[width=0.45\linewidth]{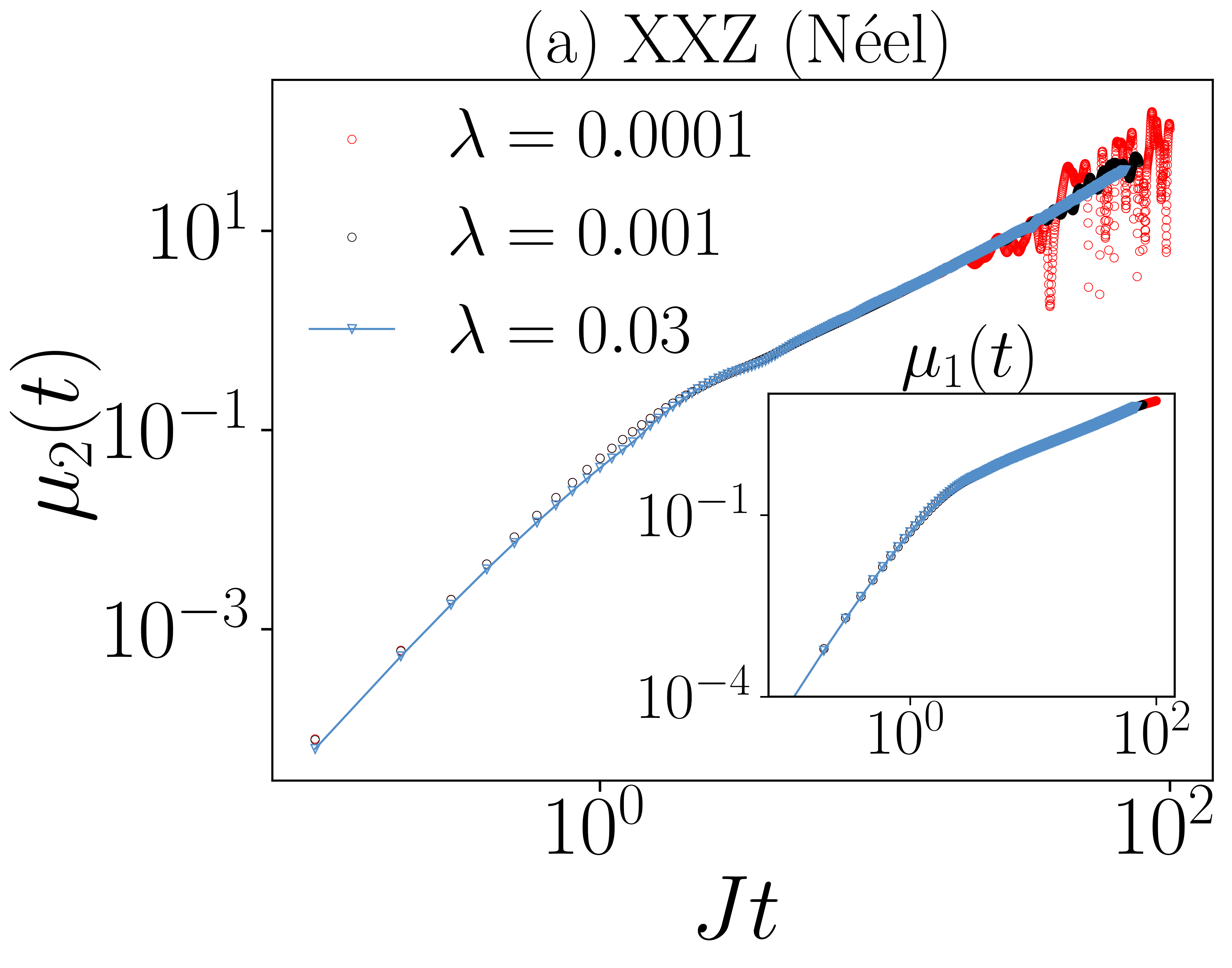}%
    \includegraphics[width=0.45\linewidth]{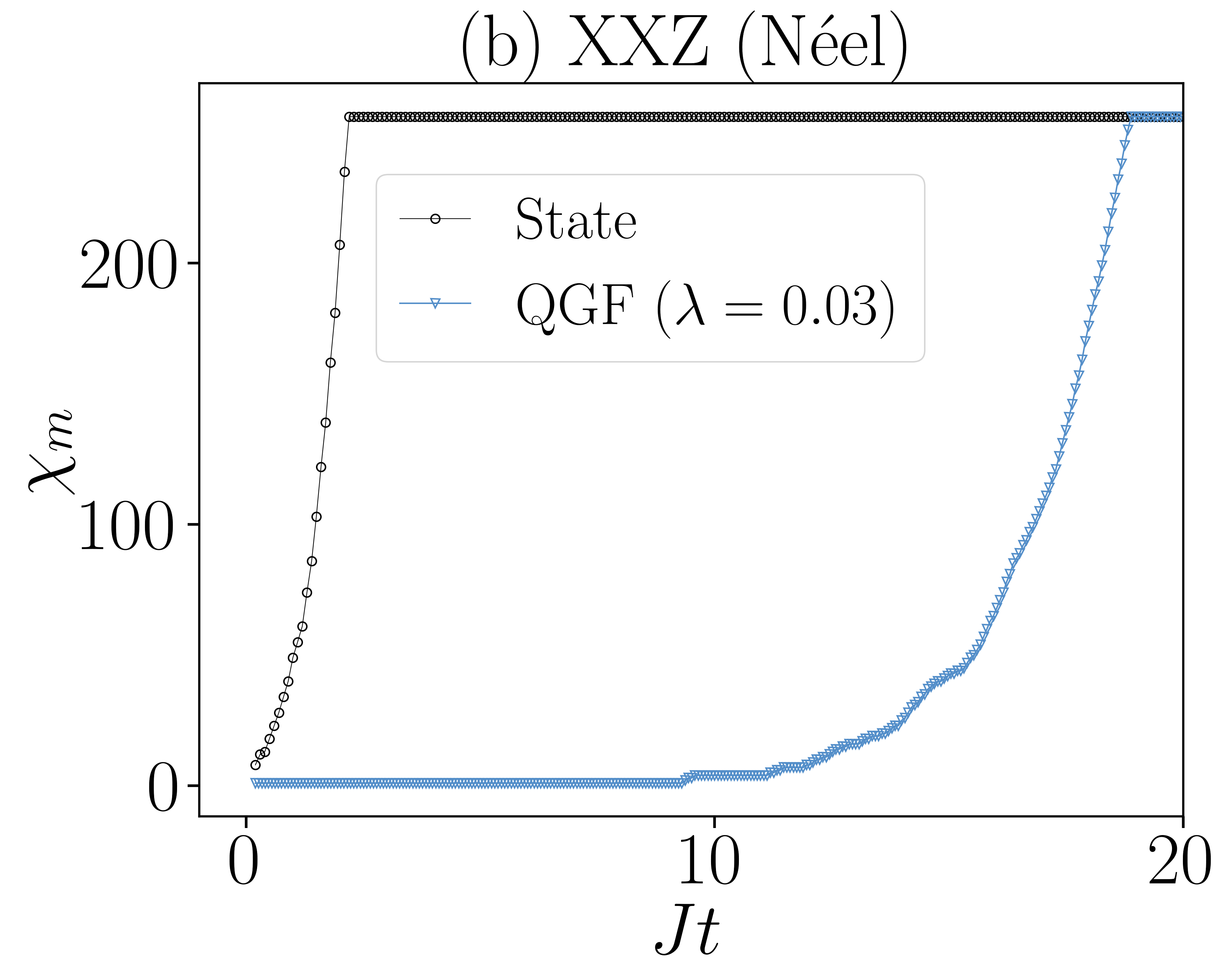}
    \caption{(a) Plot of $\mu_2(t)$ (main panel) and $\mu_1(t)$ (inset) for different values of $\lambda$, with fixed bond dimension $\chi=256$, for the XXZ chain with $\Delta=1.0J$, starting from the Néel state, for the open XXZ chain with spin excitation injection at the boundary with $\Delta=1.0J$ and $\Gamma_G=1.0J$. The Schmidt-value cutoffs are $\epsilon=10^{-10}$, $10^{-20}$, and $10^{-30}$ for $\lambda=0.03$, $0.001$, and $0.0001$, respectively. The system size is $L=40$. (b) Comparison of bond dimension growth with time between the state evolution approach and QGF approach ($\lambda=0.03$) for the middle bond $(L/2, L/2+1)$ of the lattice. The Schmidt value cutoff for both state evolution and QGF method is $\epsilon=10^{-10}$.}
    \label{fig:lamda_converge}
\end{figure}

On the other hand, choosing $\lambda$ to be very small counterintuitively makes the computation more difficult, as it requires a larger bond dimension and a much smaller Schmidt-value cutoff. This is because, for very small $\lambda$, when the counting operator approaches the identity, $R(\lambda,t)\approx\mathbb{I}$, it contains a very small $\lambda$-dependent correction. In this regime, the Schmidt values scale as $1/\lambda$, while $\mathrm{Re}[G(\lambda,t)]$ remains very close to unity up to long times. Consequently, the difference $1-\mathrm{Re}[G(\lambda,t)]$ becomes extremely small, and resolving this difference accurately requires both a larger number of Schmidt values and higher numerical precision to extract the moments reliably. Hence, choosing an excessively small $\lambda$ is also not numerically efficient.

In Fig.~\ref{fig:lamda_converge}(a), we present the first and second moments for $\lambda=0.03$, $0.001$, and $0.0001$, respectively. For all three values of $\lambda$, the long-time dynamics of the moments remain trustworthy up to $t\approx100$, which is the timescale of interest in this work. We observe that, for a fixed bond dimension $\chi=256$, $\lambda=0.03$ provides better numerical performance than the smaller values of $\lambda$. In Fig.~\ref{fig:lamda_converge}(b), we show the growth of the bond dimension at the middle bond of the lattice, which in general contains the largest bond dimension, for both the state evolution as well as the QGF method choosing $\lambda=0.03$. For the same Schmidt value cutoff, $\epsilon=10^{-10}$, we observe that the bond dimension grows much more rapidly for the state evolution than for the QGF approach. Hence, the QGF approach with an optimal $\lambda$ enables the simulation of open many-body dynamics to substantially longer times, which are otherwise inaccessible via state evolution, as also observed in Fig.~\ref{fig:setup_I} of the main text.
\begin{figure*}
    \centering
    \includegraphics[width=0.32\linewidth]{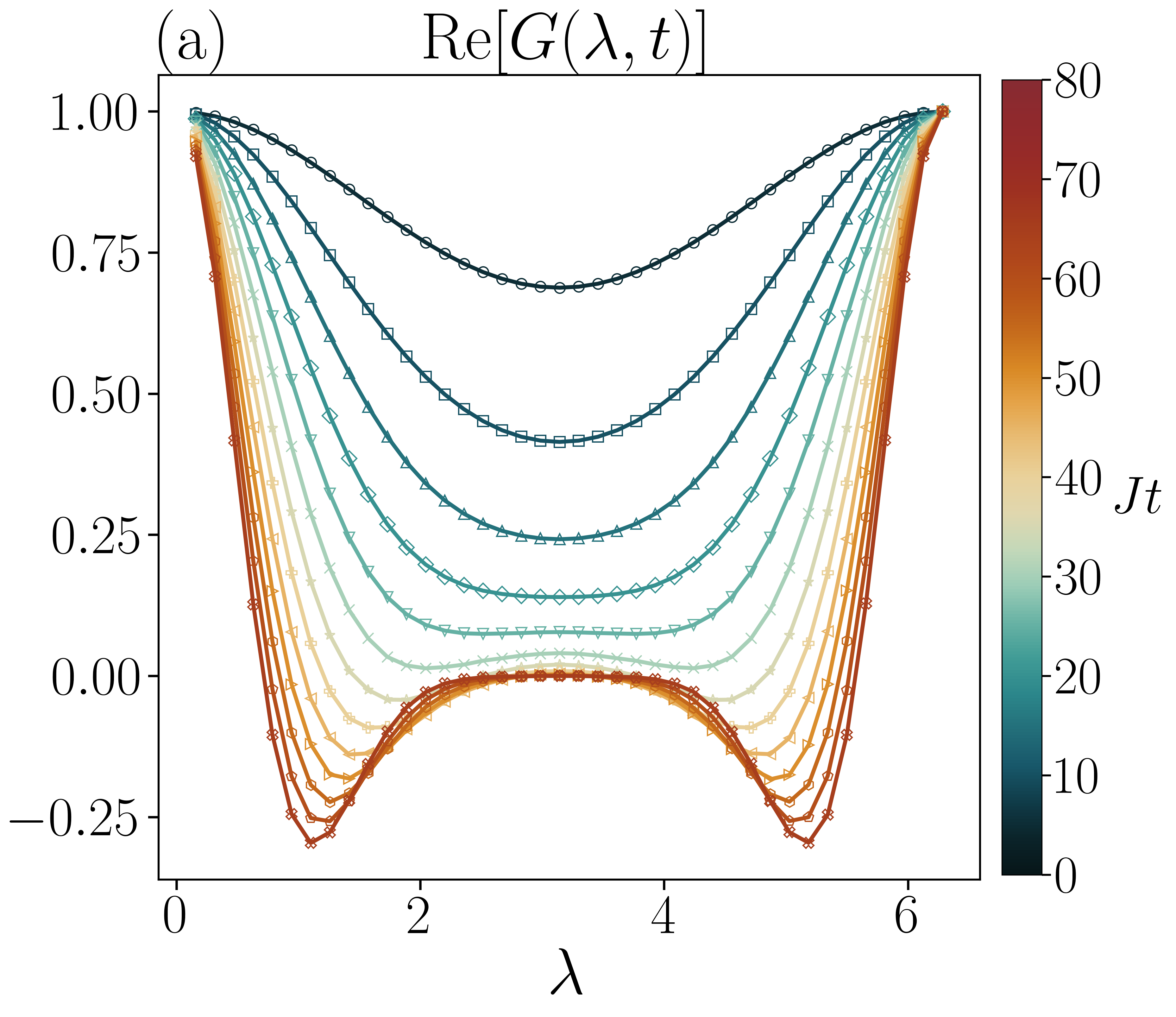}%
    \includegraphics[width=0.32\linewidth]{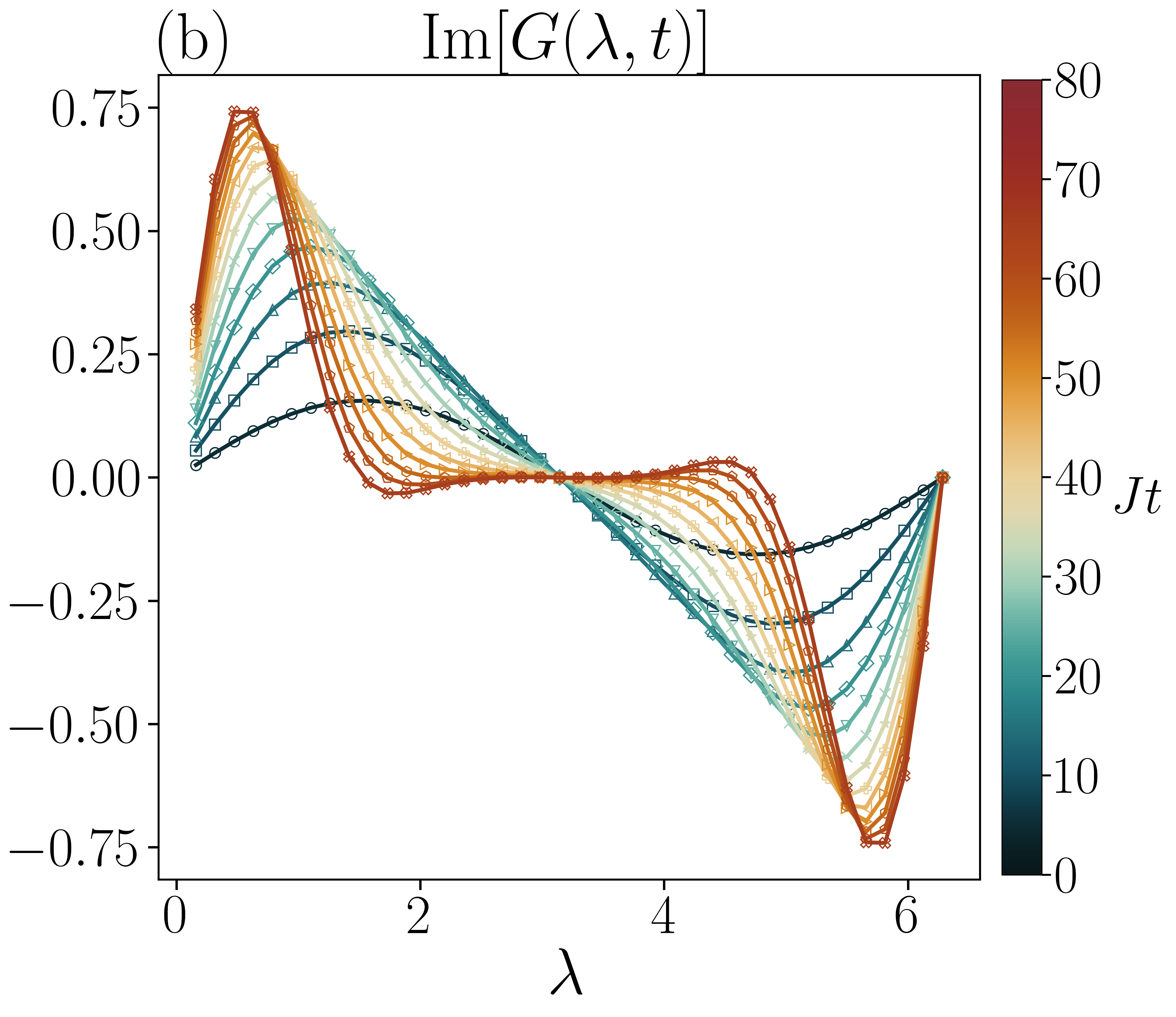}%
    \includegraphics[width=0.3\linewidth]{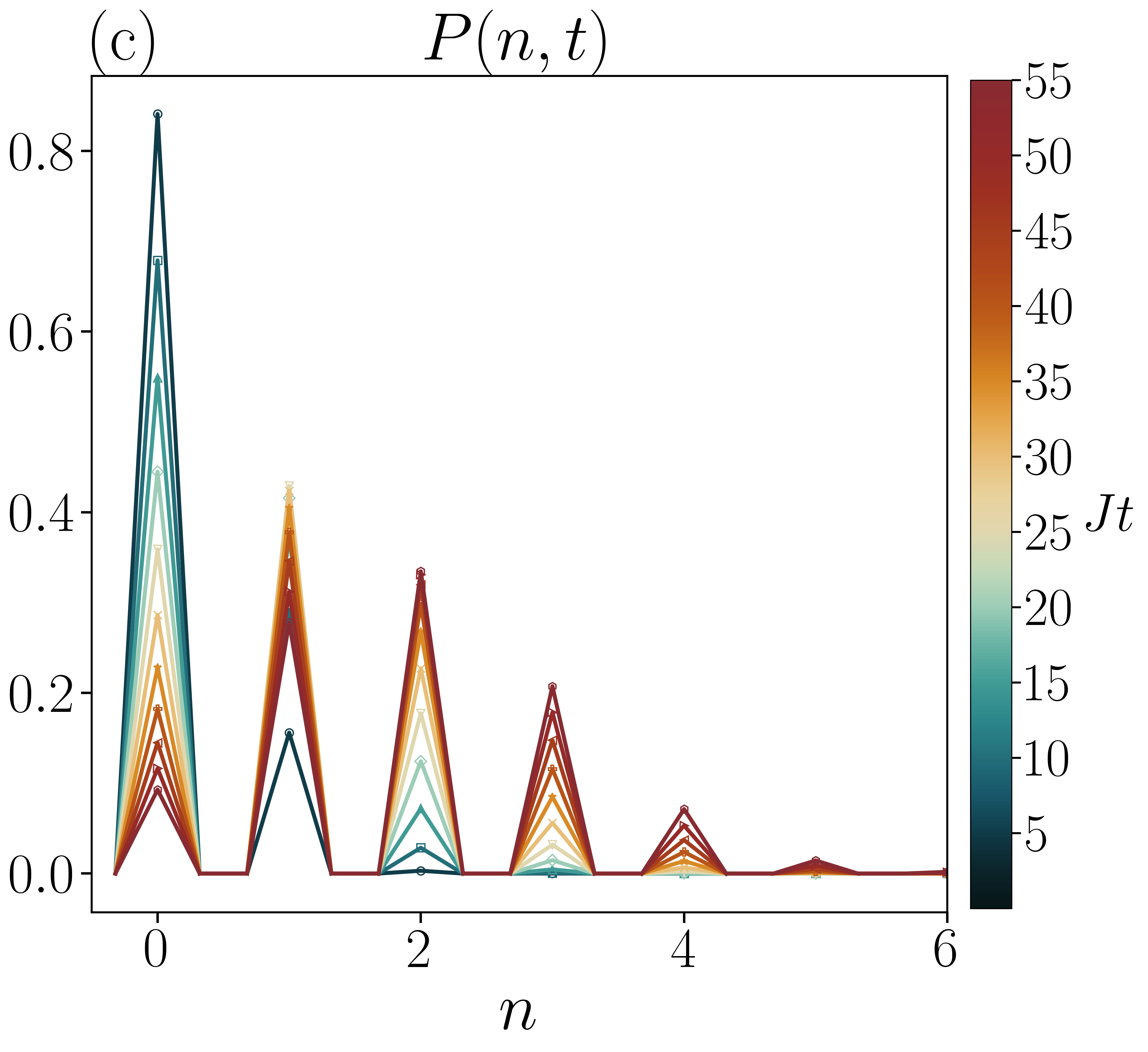}
    \caption{Plot of the real part [(a)] and imaginary part [(b)] of the quantum generating function $G(\lambda,t)$ at different time snapshots, starting from the Néel state, for the open XXZ chain with spin excitation injection at the boundary with $\Delta=1.0J$ and $\Gamma_G=0.1J$. Panel (c) shows the probability distribution $P(n,t)$ obtained by performing inverse Fourier transform of $G(\lambda,t)$. The calculations are performed using a bond dimension $\chi=512$ and a Schmidt-value cutoff $\epsilon=10^{-10}$. The system size is $L=40$. The setup of our consideration here is the lattice under spin excitation injection from the boundary.}
    \label{fig:G_lambda}
\end{figure*}
\section{Results for $G(\lambda,t)$ as a function of $\lambda$ and corresponding probability distribution}
In this section, we present the quantum generating function $G(\lambda,t)$ as a function of $\lambda$.
In Fig.~\ref{fig:G_lambda}(a) and (b), we plot both the real and imaginary part of the generating function $G(\lambda,t)$ respectively, by varying $\lambda$ between $[0,2\pi]$ at different time snapshots. In Fig.~\ref{fig:G_lambda}(c), we present the corresponding probability distribution of the injected charge by performing inverse Fourier transform $P(n,t)=\int_{0}^{2\pi} \, \frac{d\lambda}{2\pi} \,e^{i\lambda n}\, G(\lambda,t)$. 
\section{QGF approach for weakly dephased XXZ chain}
In this section, we discuss another setup where the developed QGF approach is efficient for long time dynamics. We consider XXZ and nnn-XXZ chain subjected to local dephasing at each lattice site. The corresponding GKSL equation is given by,
\begin{align}
    \frac{d\rho}{dt}=\mathcal{L}\rho=-i[\hat{H},\rho]+\sum_{j=1}^{L}\Gamma_D\Big(\hat{S}_j^z\rho \hat{S}_j^z-\frac{1}{2}\{\hat{S}^z_j \hat{S}^z_j,\rho\}\Big), \label{eq:GKSL_supp}
\end{align}
where $\Gamma_D$ is the dephasing strength.
The lattice is initialized in the N\'eel state $\rho_0=|\uparrow \downarrow\uparrow\downarrow\dots\ra\la\uparrow \downarrow\uparrow\downarrow\dots|$. Here we are interested in the statistics of the transferred magnetization within a domain of size $l$ for a lattice of length $L$ ($L\gg l)$. We choose this domain in the middle of the lattice starting from the site $i_{\rm in}=1+(L-l)/2$ to the site $i_{\rm end}=(L+l)/2$. Therefore, the magnetization operator in the domain $\hat{S}^z=\sum_{j=i_{\rm in}}^{i_{\rm end}}\hat{S}^z_j$ and the transferred magnetization operator is $\Delta \hat{S}^z = \hat{S}^z(t) - S^z(0)$. For small dephasing strength $\Gamma_D$, the evolution of the state $\rho$ using the GKSL equation is nontrivial due to the growth of operator space entanglement entropy. Hence we again evolve the counting operator $R(\lambda,0)$ under the adjoint Liouvillian superoperator to obtain $G_\phi(\lambda,t)$ and using the QGF, we obtain the moments. All the odd moments of transferred magnetization are zero for this setup. Hence, we compute the second cumulant $\kappa_2(t)$ following Eq.~\eqref{eq:mu_2} which is same as second moment. This $\kappa_2(t)$ corresponds to the fluctuation of the transferred magnetization in the domain.

\begin{figure}[h!]
    \centering
    \includegraphics[width=0.35\linewidth]{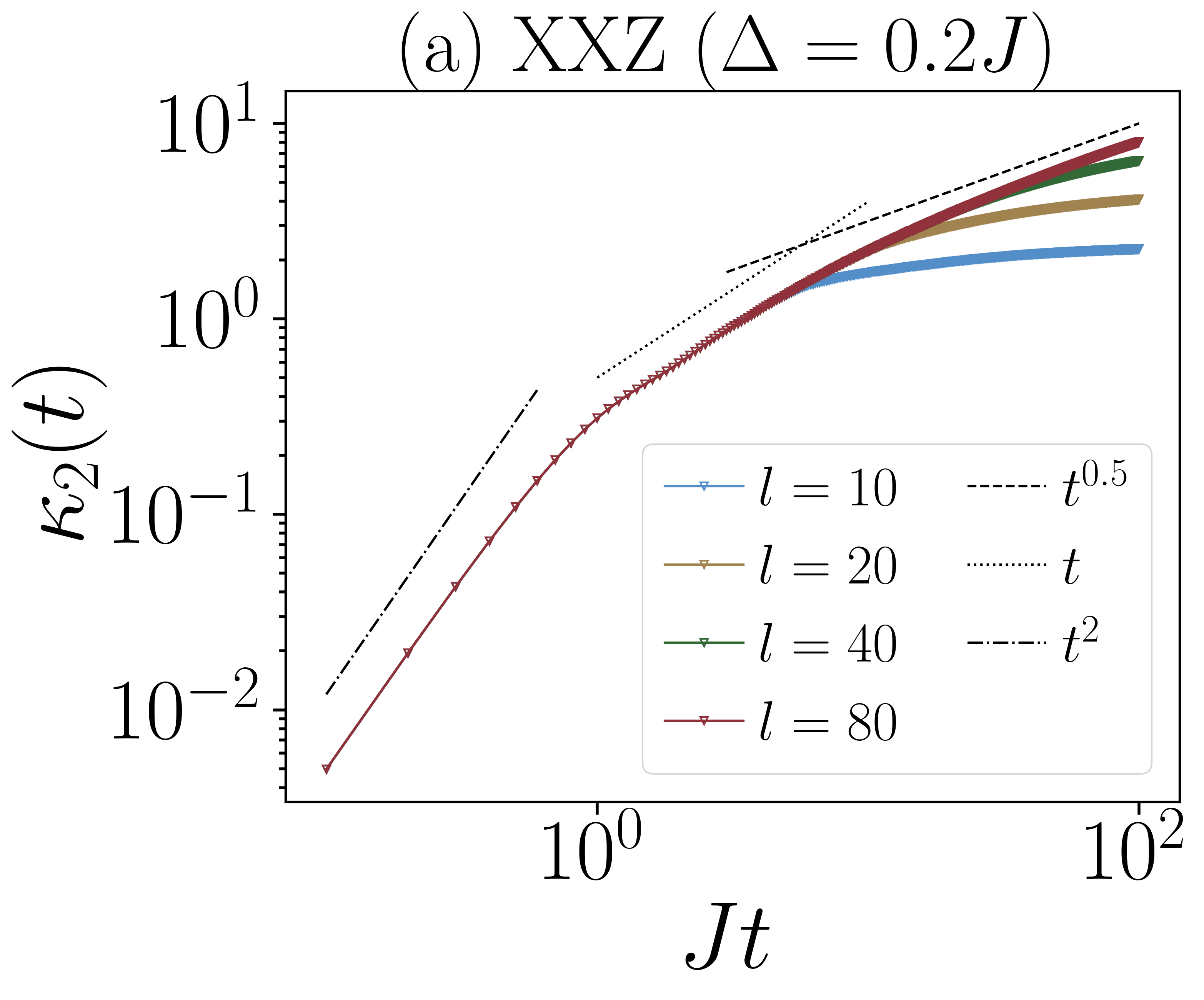}%
    \includegraphics[width=0.35\linewidth]{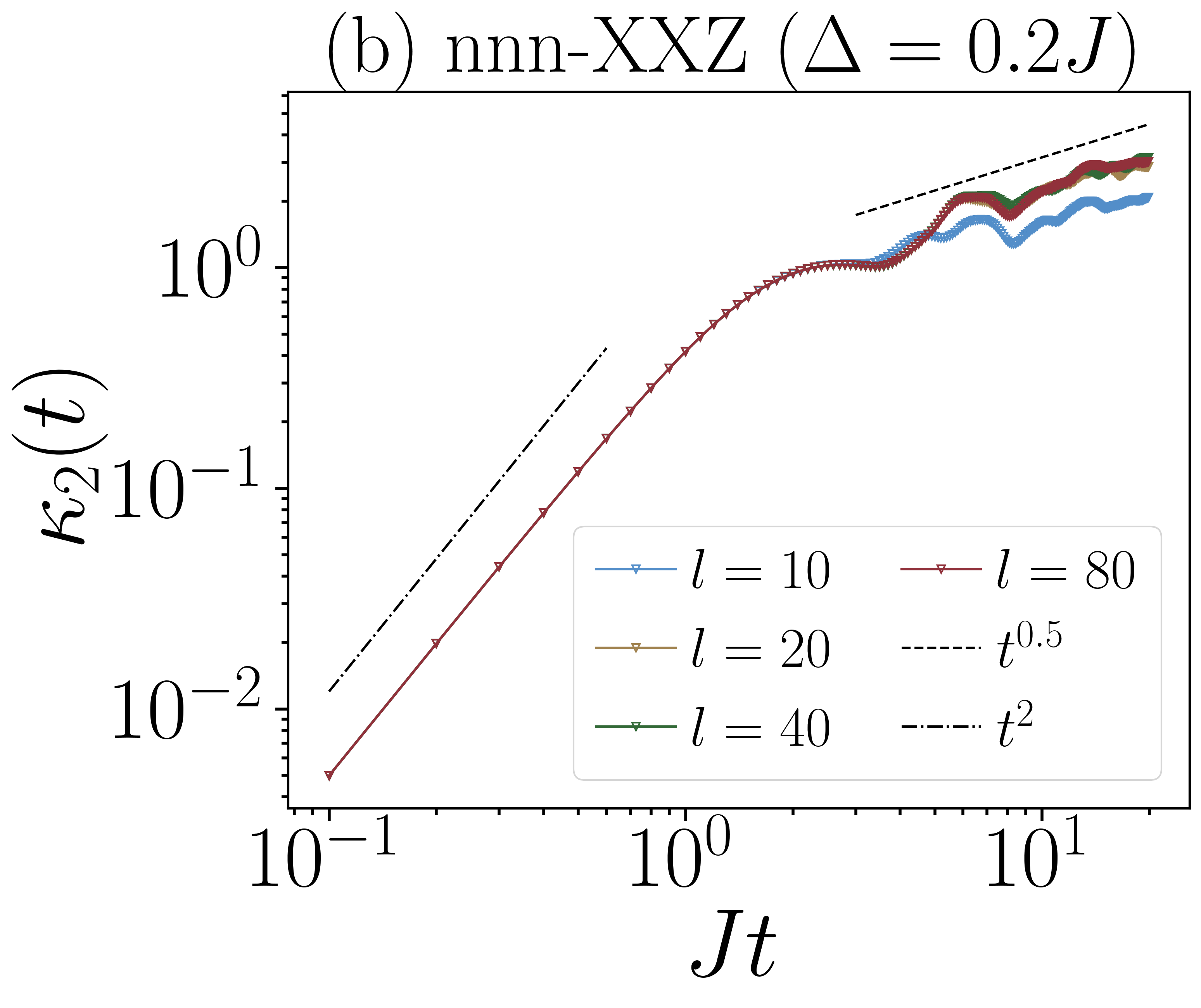}
    \caption{Plot of fluctuations of the transferred magnetization in domain of size $l$ within a domain for XXZ chain [(a)] and next-nearest-neighbour XXZ (nnn-XXZ) chain [(b)]. The size of the full system is $L=500$ and domain sizes are $l=10,20,40,80$. Dephasing strength is $\Gamma_D=0.1J$. The other parameters are $\lambda=0.03$, $\phi=0$, $J_b=J$. To simulate the dynamics, we have used TEBD algorithm with bond dimension $\chi=256$ and cutoff $10^{-10}$. }
    \label{fig:setup_II}
\end{figure}

In Fig.~\ref{fig:setup_II}, we plot fluctuations in the transfer magnetization, $\kappa_2(t)$ for domain sizes $l=10, 20, 40$, and $80$ in both the XXZ and nnn-XXZ chains with $\Delta=0.2J$. To eliminate boundary effects, the total system size is chosen to be much larger than the domain size i.e., $L=500$ in our case. The numerical implementation of the dynamics using the TEBD algorithm, as well as the choice of $\lambda$ and $\phi$, is identical to that employed in Setup-I discussed in the main text. We use a bond dimension of $\chi=256$ and cutoff $10^{-10}$. The transport classes can be identified from the dynamics of $\kappa_2(t)$ as: ballistic class correponds to $\kappa_2(t)\propto t$, diffusive class corresponds to $\kappa_2(t)\propto \sqrt{t}$, superdiffusive and subdiffusive class correspond to $\kappa_2(t) \propto t^{\nu}$ with $1/2<\nu<1$ and $\nu<1/2$ respectively.
For both models, the early-time behavior of $\kappa_2(t)$ exhibits a quadratic growth in time, $\kappa_2(t)\propto t^2$. In the XXZ chain, this is followed by a transient ballistic transport $\kappa_2(t)\propto t$ and finally diffusive dynamics $\kappa_2(t)\propto \sqrt{t}$ due to the presence of dephasing. In contrast, the nnn-XXZ chain does not display a ballistic regime beyond the initial quadratic growth and instead exhibits diffusive dynamics at later times. 
\end{document}